\documentclass[journal]{IEEEtran}
\ifCLASSINFOpdf
\else
\fi

	\usepackage[utf8]{inputenc} 
	\usepackage[T1]{fontenc}
	\usepackage{url}
	\usepackage{ifthen}
	\usepackage{cite}
	\usepackage[cmex10]{amsmath} 
	\usepackage{theorem}  
	\usepackage{graphicx}
	\usepackage{subcaption}
	\usepackage{amssymb}
	\usepackage[noend,ruled]{algorithm2e}
	\usepackage{float}
	\usepackage{enumitem}
	\usepackage{color}
	\usepackage{soul}
	\usepackage{mathtools}
	
	\theoremstyle{plain} \theorembodyfont{\normalfont\slshape}
	
	\newtheorem{thm}{Theorem$\!$}
	\newenvironment{theorem}{\begin{thm}\hspace*{-1ex}{\bf.}}{\end{thm}}
	
	\newtheorem{prop}[thm]{Proposition$\!$}
	\newenvironment{proposition}{\begin{prop}\hspace*{-1ex}{\bf.}}{\end{prop}}
	
	\newtheorem{lem}[thm]{Lemma$\!$}
	\newenvironment{lemma}{\begin{lem}\hspace*{-1ex}{\bf.}}{\end{lem}}
	
	\newtheorem{cor}[thm]{Corollary$\!$}

	\newtheorem{prob}[thm]{Problem$\!$}

	\newtheorem{cl}[thm]{Claim$\!$}

	\newtheorem{ob}[thm]{Observation$\!$}
	\newenvironment{observation}{\begin{ob}\hspace*{-1ex}{\bf.}}{\end{ob}}
	
	\newtheorem{rem}[thm]{Remark$\!$}

	\newtheorem{defi}[thm]{Definition$\!$}
	\newenvironment{definition}{\begin{defi}\hspace*{-1ex}{\bf.}}{\end{defi}}
	\newenvironment{Definition}{\begin{defi}\hspace*{-1ex}{\bf.}}{\end{defi}}

	\newcounter{consCount}
	\newenvironment{construction}{\bigskip\noindent\refstepcounter{consCount}
		\textbf{Construction \theconsCount .}}
	{\par\bigskip}
	
	\newcommand{\cC}{{\cal C}}
	\newcommand{\cD}{{\cal D}}
	\newcommand{\cE}{{\cal E}}
	\newcommand{\cF}{{\cal F}}

	\newcommand{\cI}{{\cal I}}
	\newcommand{\cJ}{{\cal J}}

	\newcommand{\cP}{{\cal P}}
	\newcommand{\cQ}{{\cal Q}}
	\newcommand{\cR}{{\cal R}}

	\newcommand{\cV}{{\cal V}}

	\newcommand{\cY}{{\cal Y}}

	\newcommand{\bfa}{{\boldsymbol a}}
	\newcommand{\bfb}{{\boldsymbol b}}
	\newcommand{\bfc}{{\boldsymbol c}}
	
	\newcommand{\bfe}{{\boldsymbol e}}
	
	\newcommand{\bfg}{{\boldsymbol g}}

	\newcommand{\bfs}{{\boldsymbol s}}
	
	\newcommand{\bfu}{{\boldsymbol u}}
	\newcommand{\bfv}{{\boldsymbol v}}
	\newcommand{\bfw}{{\boldsymbol w}}
	\newcommand{\bfx}{{\boldsymbol x}}
	\newcommand{\bfy}{{\boldsymbol y}}
	\newcommand{\bfz}{{\boldsymbol z}}

	\newcommand{\bfH}{{\mathbf H}}

	\newcommand{\bfS}{{\mathbf S}}

	\newcommand{\bfX}{{\mathbf X}}
	\newcommand{\bfY}{{\mathbf Y}}
	
	\newcommand{\bfSig}{{\mathbf{\Sigma}}}

	\newcommand{\Ppnexc}{P_{\mathsf{suc}}^{(\mathsf{p})}}

	\newcommand{\Ppmiv}{P_{\mathsf{miv}}^{(\mathsf{p})}}

	\newcommand{\Prmiv}{P_{\mathsf{r.m}}}
	\newcommand{\cPrmiv}{\cP_{\mathsf{r.m}}}

	\newcommand{\Ppfail}{P_{\mathsf{fail}}^{(\mathsf{p})}}

	\newcommand{\Pmgz}{P_{\mathsf{miv}\scriptscriptstyle{>}\scriptstyle{0}}^{\mathsf{(i.p)}}}

	\newcommand{\ners}{{m_1}}
	\newcommand{\nerr}{{m_2}}

	\newcommand{\sd}{\mathsf{d}}
	\newcommand{\sdoc}{\mathsf{d_{s.c}}}
	\newcommand{\sdeloc}{{\delta_\mathsf{s.c}}}
	\newcommand{\sdelocr}{{\delta_\mathsf{s.c}^{(r)}}}
	
	\newcommand{\sdeloct}{{\delta'_\mathsf{s.c}}}
	\newcommand{\doc}{d_\mathsf{s.c}}
	
	\newcommand{\bSig}{\mathbf{\Sigma}}
	
	\newcommand{\isd}{{i_{\mathsf{d}}}}
	
	\newcommand{\Pers}{P_{\mathsf{ers}}}
	\newcommand{\Perr}{P_{\mathsf{err}}}

	\newcommand{\phit}{\tilde{\phi}_1}

	\newcommand{\cQpas}{\cQ_\mathsf{a.s}^\mathsf{(p)}}
	\newcommand{\cQias}{\cQ_\mathsf{a.s}^\mathsf{(i.p)}}
	\newcommand{\alioff}{w}
	\newcommand{\dify}{\bfz}
	
	\DeclareMathOperator*{\argmax}{arg\,max}
	\DeclareMathOperator*{\argmin}{arg\,min}
	
	\newcommand{\binsynd}{{\boldsymbol \sigma}}
	\newcommand{\algsynd}{\underline{\boldsymbol \sigma}}

	\SetCommentSty{mycommfont}

	\definecolor{brown}{rgb}{0.8,0.7,0.55}
	
	\definecolor{purple}{rgb}{0.5,0.0,0.5}

\begin{document}
		%
		\title{Genomic Compression with Read Alignment at the Decoder}
		%
		%
		%
		
		\author{Yotam~Gershon,~\IEEEmembership{Student Member,~IEEE,} and
			Yuval~Cassuto,~\IEEEmembership{Senior Member,~IEEE,}
			\thanks{Y. Gershon and Y. Cassuto are with the Viterbi Department of Electrical and Computer Engineering, Technion - Israel Institute of Technology. This work was supported in part by the Israel Science Foundation under grant number 2525/19.
				
				The results in this paper were presented in part at the 2021 and 2022 International Symposiums on Information Theory.}}

\maketitle
\begin{abstract}
	We propose a new compression scheme for genomic data given as sequence fragments called reads. The scheme uses a reference genome at the decoder side only, freeing the encoder from the burdens of storing references and performing computationally costly alignment operations. The main ingredient of the scheme is a multi-layer code construction, delivering to the decoder sufficient information to align the reads, correct their differences from the reference, validate their reconstruction, and correct reconstruction errors. The core of the method is the well-known concept of distributed source coding with decoder side information, fortified by a generalized-concatenation code construction enabling efficient embedding of all the information needed for reliable reconstruction. We first present the scheme for the case of substitution errors only between the reads and the reference, and then extend it to support reads with a single deletion and multiple substitutions. A central tool in this extension is a new distance metric that is shown analytically to improve alignment performance over existing distance metrics.
\end{abstract}


%
\IEEEpeerreviewmaketitle

\vspace*{-2ex}

\section{Introduction}
%
%
%
%
\IEEEPARstart{G}{enomic sequencing} is the process of analyzing the order of nucleic acids within a DNA molecule (genomic sequence). In many modern sequencing technologies, a large set of short sequence fragments, called \emph{reads}, is produced and represented by a string of characters (usually \textit{A,C,G,T}). In this method, called \emph{shotgun sequencing} \cite{sps5}, each read's location within the sequence is generally unknown. Furthermore, the sequencing machine introduces mutation errors into the reads. Therefore, the sequence assembly from the reads requires a large number of reads and high computational effort. Effective compression of pre-assembly reads data is therefore an essential problem. A large number of read-compression methods are available \cite{sps6, sps9, sps21}, partitioned into two main categories: reference-free and reference-based tools, differing by whether a closely similar reference genome is shared by the encoder and decoder. In some applications, genome reads will be produced at an edge node, e.g. a physician's office, and then sent for processing to a central node, e.g. a cloud database. In such scenarios, the cost of known reference-based and reference-free methods may be too high for the edge node's limited resources, due to the need to store long references and perform costly alignment (in the former), and cluster reads and use powerful compressors (in the latter). Alternatively, in this paper we propose a compression scheme in which the encoder needs to neither store references nor perform heavy processing, while benefiting from a reference available at the central node decoding the reads. 

Compression with a reference, acting as side information available only at the decoder, is an instance of the well known Slepian-Wolf (SW) coding problem~\cite{sps2}. This source coding problem was first described as an equivalent channel coding problem over a virtual difference channel between the sources by Wyner~\cite{sps30}, and an explicit construction based on the syndromes of error-correcting codes was first introduced in the pioneering work of Pradhan and Ramchandran~\cite{sps10}. Several works offered improved constructions and/or analysis approaches~\cite{sps11,sps18,sps19,sps25,sps24,sps1}. Some works even dealt specifically with genomic data~\cite{sps22,sps23}, but focused on either full-block or streamlined data, and not fragmented reads from randomly chosen and unknown locations. When compressing fragmented reads, the known capability of reconstructing a read from a similar reference is not sufficient, and additional information is needed for aligning the read within the full reference. 

In this work, our objective is to compress a batch of reads taken from a long genome sequence, such that a decoder with a reference similar to this genome can perfectly reconstruct all the reads without error. The construction proposed in Section~\ref{Sec:CodeConst} of this paper offers both alignment and reconstruction  capabilities, using the framework of generalized error-locating (GEL) \cite{sps17} \emph{coset codes}. GEL coset codes were used in the Cassuto-Ziv construction~\cite{sps1} for full-block compression, and the novelty of this present work is in using the GEL's hierarchy of inner codes to combine all the information for read reconstruction into the same codeword: one layer for alignment, one for similarity reconstruction and one for reconstruction validation. In the fourth layer, a batch of reads is encoded with an outer code to provide extremely low failure probability. We begin by assuming that the errors between the sequence, its reads and the reference are modeled by substitution errors alone, and propose an efficient coding scheme for such a scenario. The construction uses general linear codes, and in addition we show how to specialize it using known algebraic codes such as BCH codes. 

We continue by extending the scheme to deal with a single deletion introduced to a read in addition to the substitution errors. This model is justified by substitution errors being the dominant error type in many sequencing technologies, while having some deletion errors that are more rare. Toward that extension, we propose in Section~\ref{Sec:DelSemiMetric} a new efficiently computable distance measure, called shift-compensating distance, that is analytically shown to reduce the false-alignment probability significantly compared to a commonly-used distance measure. Then, in Section~\ref{Sec:SchemeDelExt}, we modify our coding scheme to support these deletion errors. In Section~\ref{Sec:SchemeAnalysis}, we pursue the probabilistic analysis of the scheme. These results are used as a design tool for setting the scheme's parameters for compression rate minimization, while meeting a given specification of perfect read-recovery success probability. 
Finally, in Section~\ref{Sec:Future} we suggest further extensions, and in Section~\ref{Sec:conclude} we colnclude and suggest topics for future work.

\begin{figure*}[ht!]
	\centering
	\begin{subfigure}{.5\textwidth}
		\centering
		\includegraphics[width=0.93\linewidth]{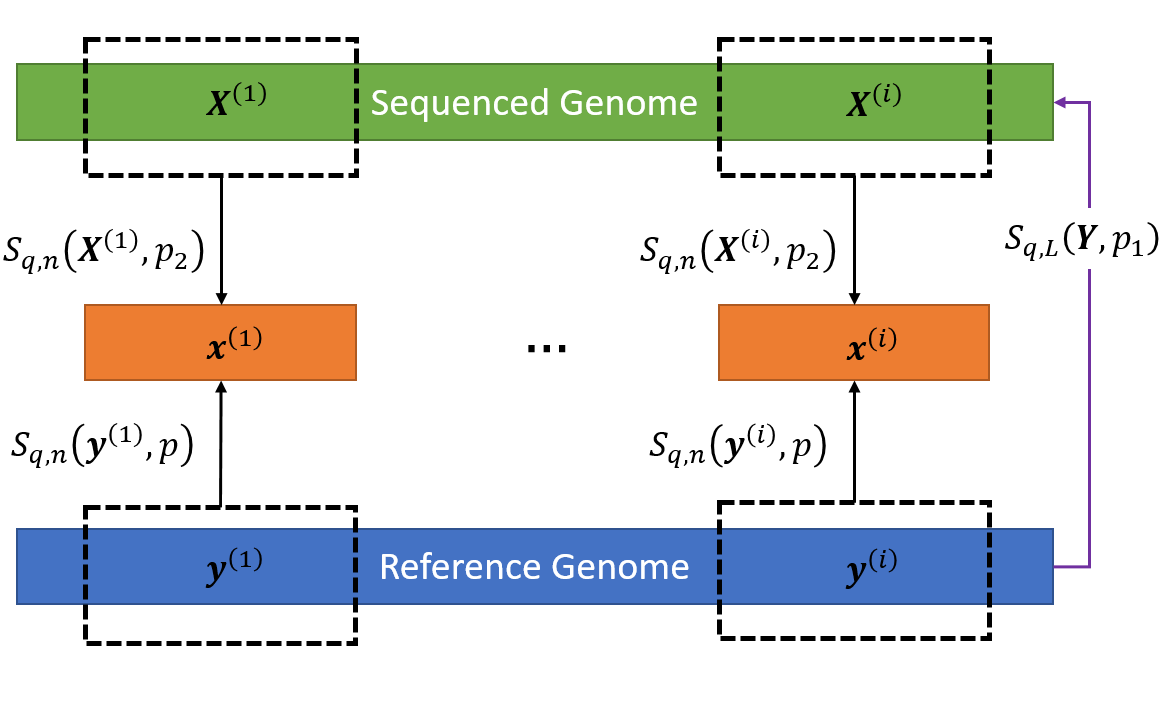}
	\end{subfigure}%
	\begin{subfigure}{.5\textwidth}
		\centering
		\includegraphics[width=0.85\linewidth]{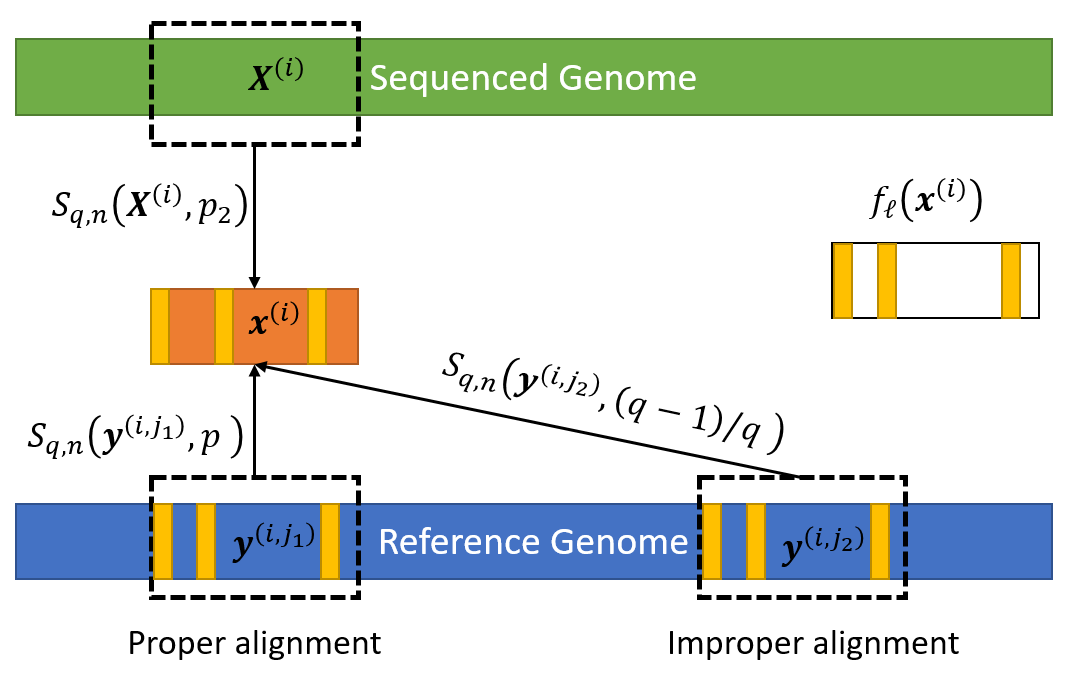}
	\end{subfigure}
	
	\caption{\textbf{Left:} DNA reads generated as fragments of a genome. \textbf{Right:} read-identifier-based alignment of a single read.}
	\label{fig:DNAsetup}
	
\end{figure*}

\section{Problem Formulation}\label{Sec:ProbForm}
\subsection{Problem Setting}
A genome data being sequenced in an edge node is represented by a string $\bfX \in \bfSig^L$, for some alphabet $\bfSig$ of size $q$. A closely similar reference of this data, represented by $\bfY\in \bfSig^L$, is stored in a central node, and is unavailable at the edge node. The differences between $\bfX$ and $\bfY$ are assumed to be caused by genetic diversity \cite{sps13}. A sequencing machine at the edge node generates a batch of $M$ \emph{reads} from $\bfX$, denoted by $\{\bfx^{(i)}\}_{i=1}^{M}$. Each read $\bfx^{(i)} \in \bfSig^n$ is an \emph{approximate sub-sequence} of $\bfX$, taken from an unknown random location within $\bfX$, and introduced with \emph{sequencing errors}. Our goal is to encode the reads $\{\bfx^{(i)}\}_{i=1}^{M}$ to a minimal size, such that they can be perfectly recovered from the similar reference $\bfY$. The encoder output is communicated to the central node \emph{without errors}.

\subsection{Substitutions Channel Model}
\label{SubSec:SubsChannel}
We begin by modeling errors using a substitutions-only channel; later in the paper we also address deletion errors. The \emph{genetic diversity} is modeled as $\bfX = \mathsf{S}_{q,L}(\bfY,p_1)$, a $q$-ary symmetric channel for which
\begin{align*}
	\forall j \in \{1,\dots,L\} : P(X_j|Y_j) = \begin{aligned}
		\begin{cases}
			1-p_1 & X_j = Y_j \\ p_1/(q-1) & X_j \neq Y_j
		\end{cases}
	\end{aligned} .
\end{align*}
We now let $\bfX^{(i)} = X_{r_i},X_{r_i+1},\dots,X_{r_i+n-1}$ be a sub-sequence observed by the sequencer, where $r_i$ is a random, unknown \emph{starting index}. A read $\bfx^{(i)}$ with \emph{sequencing errors} is modeled by a similar $q$-ary symmetric channel: $\bfx^{(i)} = \mathsf{S}_{q,n}(\bfX^{(i)},p_2)$. 
We call $\bfy^{(i)} = Y_{r_i},Y_{r_i+1},\dots,Y_{r_i+n-1}$ the sub-sequence of $\bfY$ with \emph{proper alignment} to the read $\bfx^{(i)}$. 

\begin{lemma}
	\label{Lem:EquivChan}
	Let $p \triangleq p_1\cdot \left[1-p_2/(q-1)\right] + (1-p_1) \cdot p_2$. Then for a proper alignment we have $\bfx^{(i)} = \mathsf{S}_{q,n}(\bfy^{(i)},p) $.
\end{lemma}

\begin{IEEEproof}
	It follows that
	\begin{align*}
		P & (y_j\neq x_j) \\
		& = P(Y_{r_i+j} \neq X_{r_i+j}) [1-P(x_j \neq X_{r_i+j} \wedge x_j = Y_{r_i+j})] \\
		& \qquad + P(Y_{r_i+j} = X_{r_i+j})\cdot [1-P(x_j = X_{r_i+j})] \\ 
		& = p_1\cdot \left[1-p_2/(q-1)\right] + (1-p_1) \cdot p_2 = p .
	\end{align*}
\end{IEEEproof}

The relations between the genome, reference and reads are illustrated in Fig.~\ref{fig:DNAsetup} (left). 

\textit{Remark:} Although in the statistical model in Lemma~\ref{Lem:EquivChan} we define $\bfy^{(i)}$ as the channel \emph{input}, one should notice that in the operational problem we need to reconstruct $\bfx^{(i)}$ from $\bfy^{(i)}$, and not the other way around as in standard channel coding.

The induced equivalent channel serves as the \emph{difference channel} between the side information (reference segment) and the source (corresponding read), for which an error-correcting code should be designed~\cite{sps30,sps10}. In our coding scheme, we will add a layer enabling the alignment of each read to the reference in order to find a closely matching segment to reconstruct from.
A modification of the scheme to deal with a channel model including a single deletion per read in addition to the substitution errors is developed in Sections~\ref{Sec:DelSemiMetric},\ref{Sec:SchemeDelExt}.
We choose in this paper the channel model with a single parameter $p$ for simplicity, while in future work one may consider richer models (e.g. \cite{sps8}), potentially exploiting the dependence between two overlapping reads, whose references pass through the same instantiation of the channel $\mathsf{S}_{q,L}(\bfY,p_1)$.

\subsection{Pre-Decoding Alignment}
\label{SubSec:Alignment}
We wish to encode $\bfx^{(i)}$ such that only the information required to reconstruct it from $\bfy^{(i)}$ is transmitted. Nevertheless, the decoder first needs to know the starting index $r_i$. Therefore, the encoder must transmit some additional information enabling the decoder to align the read within the reference, while accounting for the substitution errors. This alignment information is an $\ell$-bit function $f_\ell(\bfx^{(i)})$ called \emph{read identifier}. The value of $\ell$ is set as a tradeoff between alignment quality and compression rate. Since only partial information is provided for alignment, additional \emph{improper alignments}, i.e., erroneous starting indices, are likely to be found. The alignment process provides the decoder with a set $\left\lbrace\bar{\bfy}^{(i,j)} | j=1,2,\dots \right\rbrace$ of length-$n$ segments of $\bfY$ as candidates for read reconstruction. Every $\bar{\bfy}^{(i,j)}\neq\bfy^{(i)}$ can be regarded as a sequence from which $\bfx^{(i)}$ is obtained by passing through a useless channel $\mathsf{S}_{q,n}(\cdot,(q-1)/q)$ with zero mutual information. The alignment process is illustrated in Fig.~\ref{fig:DNAsetup} (right). Clearly, only the proper alignment is desired for decoding, thus a method for rejecting improper candidates is required. This method, described in Section~\ref{Sec:CodeConst}, is referred to as \emph{validation}.

\section{The Compression Scheme}
\label{Sec:CodeConst}

For simplicity, from this point on we assume $q=2$, i.e. $\bfSig = \{0,1\}$, but every result can be generalized to any  $q$ that is a prime power. Furthermore, cyclic 1-based indices will be used, i.e., every index $j$ in $\bfX,\bfY$ will be taken as $j \cong \left[(j-1) \mod L\right] + 1  \in \{1,\dots,L\}$ to avoid edge effects and for the ease of notation.

\vspace*{-1.5ex}
\subsection{Read Identifier}
A simple \emph{bit sampling} approach is found to be very suitable for read identifiers. Let $1 \leq i_1 < \dots < i_\ell \leq n$ be a predefined set of indices, known to both the encoder and decoder. Now, let $f_\ell(\bfx^{(i)}) = x_{i_1},\dots,x_{i_\ell}$ be the read identifier. In this case, an \emph{aligner} at the central node simply correlates this identifier along the reference using the Hamming distance with respect to each starting index, and produces the set
\begin{equation}
	\label{Eq:Candidates}
	\mathrm{Y}^{(i)} = \left\lbrace \left. \bar{\bfy}^{(i,j)} \right| d_H\left(f_\ell(\bfx^{(i)}) , f_{\ell}(\bar{\bfy}^{(i,j)}) \right) \leq \mathsf{T} \right\rbrace_{j=1}^{\mathsf{K}_i},
\end{equation}
where $d_{\mathsf{H}}(\cdot,\cdot)$ is the Hamming distance, $\mathsf{T}$ is a predefined threshold, and 
\begin{equation*}
	\left\lbrace \bar{\bfy}^{(i,j)} = \left[Y_{r_i^{(j)}},\dots,Y_{r_i^{(j)}+n-1}\right] \right\rbrace
\end{equation*}
is the set of $\mathsf{K}_i$ \emph{matching} alignments of $\bfx^{(i)}$ within $\bfY$. The remainder of the read, i.e., the indices outside of the identifier, is denoted by $\bfx^{(i)}_{\cI}$, where $\cI=\{1,\dots,n\}\setminus\{i_1,\dots,i_\ell\} , |\cI|=n-\ell$. In Section~\ref{SubSec:SubCandMod} we will show that taking $\{i_1,\dots,i_\ell\}$ evenly spaced over the read is preferable over a simple prefix when the model includes deletion differences.

\vspace*{-2ex}
\subsection{General Code Construction}
\label{subsec:coding_scheme}
We wish to encode and transmit a batch of reads $\{\bfx^{(i)}\}_{i=1}^{M}$ from $\bfX$ such that a decoder with access to $\bfY$ satisfying $ \bfX = \mathsf{S}_{2,L}(\bfY,p_1)$ will perfectly reconstruct them with high probability. Recall from Section~\ref{SubSec:SubsChannel} that $p$ is the equivalent error probability between a proper alignment $\bfy^{(i)}$ in $\bfY$ to its corresponding read $\bfx^{(i)}$.

\begin{definition}
	A $\mathbf(M,n,\cR,p,P_\mathsf{s})$-code is a pair $(\cE,\cD)$ of encoder-decoder for a set $\{\bfx^{(i)}\}_{i=1}^{M}$ of length-$n$ reads such that:
	\begin{enumerate}
		\item $\cE$ has access only to $\{\bfx^{(i)}\}_{i=1}^{M}$ ,
		\item $\cD $ has access only to $\bfY$ and $\cE\left(\{\bfx^{(i)}\}_{i=1}^{M}\right)$,
		\item the encoded size satisfies $\left|\cE\left(\{\bfx^{(i)}\}_{i=1}^{M}\right)\right| = (nM) \cdot \cR$,
		\item the correct-decoding probability satisfies
		\begin{equation*}
			\Pr \left\lbrace \cD \left[ \cE\left(\{\bfx^{(i)}\}_{i=1}^{M}\right) , \bfY \right] = \{\bfx^{(i)}\}_{i=1}^{M} \right\rbrace \geq P_\mathsf{s}.
		\end{equation*}
	\end{enumerate}
\end{definition}
$P_\mathsf{s}$ is a decoding {\em success probability} requirement specified for the coding scheme, and $\cR$ is its (fixed) compression rate. Our general code construction is based on generalized error locating (GEL) codes \cite{sps17}, and specifically on the Cassuto-Ziv code construction~\cite{sps1}, adapted to use as a source code with alignment-validation capabilities.

\begin{construction}
	\label{cons:GenCons}
	Let $\cC_1,\cC_2$ be a pair of binary linear codes with parameters $\left[n-\ell,k_j-\ell,d_j\right], {j=\{1,2\}}$, where $k_1 \geq k_2$. Let  $\mathsf{H}_1,\mathsf{H}_2$ be parity-check matrices of these codes, respectively, such that $\mathsf{H}_2$ is obtained from $\mathsf{H}_1$ by concatenating a \emph{validation matrix} $\bar{\mathsf{H}}_2$ of $\tau \triangleq k_1-k_2$ rows that are linearly independent of the rows of $\mathsf{H}_1$. Let $\mathsf{H}_{\mathsf{c}}$ be a \emph{complementary matrix} such that concatenating it with $\mathsf{H}_2$ forms a square \emph{full-rank} matrix $\mathsf{H}$. This structure is illustrated in Fig. \ref{fig:ParStructure}.
	
	\begin{figure}[H]
		\vspace*{-3ex}
		\centering
		\includegraphics[width=0.65\linewidth]{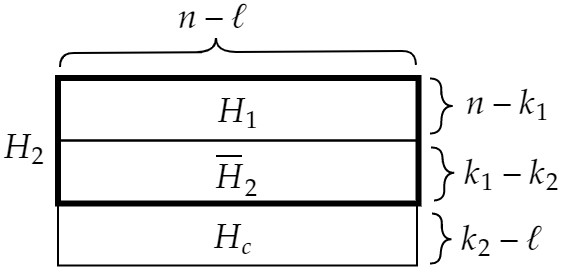}
		\caption{Structure and sizes of the construction's inner-code parity-check matrices.}
		\label{fig:ParStructure}
	\end{figure}
	
	Finally, let $\cC_\mathsf{o}$ be a $[M,k_{\mathsf{o}},d_{\mathsf{o}}]$ linear, maximum-distance separable (MDS) \cite{sps27} \emph{outer code} over $\mathsf{GF}(2^\nu)$, with parity-check matrix $\bfH_{\mathsf{o}}$, where $\nu = k_2 - \ell$. 
	
	The encoding process is given formally in Algorithm~\ref{Alg:Cons1Enc}, and we now describe its main steps in words. For every read $\bfx^{(i)}, 1\leq i\leq M$, we first extract its read identifier $\bfw^{(i)}  \triangleq f_\ell(\bfx^{(i)})$, and then for the remaining indices $\bfx_{\cI}^{(i)}$ we calculate the syndromes $\bfs^{(i)},\bfa^{(i)}$ with respect to $\mathsf{H}_2 , \mathsf{H_c}$, respectively. This process is referred to as the \emph{inner encoding}. The syndromes $\bfa^{(i)}$ are regarded as elements in $\mathsf{GF}(\nu)$ and collected to a word $\underline{\bfa} \triangleq [\bfa^{(1)},\dots,\bfa^{(M)}]$, and the syndrome $\bfS$ of $\underline{\bfa}$ with respect to $\mathbf{H}_\mathsf{o}$ is calculated. This is referred to as the \emph{outer encoding}. The encoder outputs the read identifiers $\{\bfw^{(i)}\}_{i=1}^M$, the \emph{inner syndromes} $\{\bfs^{(i)}\}$, and the \emph{outer syndrome} $\bfS$. We note that the inner syndrome can be described as $\bfs^{(i)} = [\bfs^{(i)}_1,\bfs^{(i)}_2]$, where $\bfs^{(i)}_1, \bfs^{(i)}_2$ correspond to $\mathsf{H}_1,\bar{\mathsf{H}}_2$, respectively.
	
	
	\begin{algorithm}[ht!]
		\textbf{Input}: $\{\bfx^{(i)}\}_{i=1}^{M}, \mathsf{H}_2,\mathsf{H}_\mathsf{c}, \bfH_{\mathsf{o}}$\\
		\For(\tcp*[h]{Inner Encoding}){$1\leq i \leq M$}{
			Extract $\bfw^{(i)} = f_{\ell}(\bfx^{(i)})$\\
			Calculate $\bfs^{(i)} = \mathsf{H}_2 \left[\bfx^{(i)}_{\cI}\right]^T$, $\bfa^{(i)} = \mathsf{H}_\mathsf{c} \left[\bfx^{(i)}_{\cI}\right]^T$\\
		}
		Form $\underline{\bfa} = \left[\bfa^{(1)},\dots,\bfa^{(M)}\right] \in \left[\mathsf{GF}(2^\nu)\right]^{M}$\\
		Calculate $\bfS=\bfH_{\mathsf{o}} \underline{\bfa}^T$ \tcp*[h]{Outer Encoding}\\
		\textbf{Output}:
			$\cE\left(\{\bfx^{(i)}\}_{i=1}^M\right) = \left\lbrace \{\bfw^{(i)}\}_{i=1}^{M},\{\bfs^{(i)}\}_{i=1}^{M},\bfS \right\rbrace$
		\caption{Construction \ref{cons:GenCons} Encoding}
		\label{Alg:Cons1Enc}
	\end{algorithm}
	
	The decoder for the construction is presented formally in Algorithm~\ref{Alg:Cons1Dec}, now also explained in words. Each read identifier is first \emph{aligned} over $\bfY$ to form $\mathrm{Y}^{(i)}$ from~\eqref{Eq:Candidates}. We set a flag `found' to $0$, and then for every candidate $\bfy^{(i,j)}$ we extract $\bfy^{(i,j)}_{\cI}$, and decode it with the function $\cD_1(\cdot,\cdot)$ that finds the nearest word to $\bfy^{(i,j)}_{\cI}$ with syndrome $\bfs_1^{(i)}$ with respect to $\mathsf{H}_1$. The result $\bfv$ of each decoding is \emph{validated} by computing its syndrome $\hat{\bfs}_2^{(i)}$ with respect to $\bar{\mathsf{H}}_2$, and comparing it to the received $\bfs_2^{(i)}$. If equal, and the flag `found' is $0$ (meaning it is the first candidate to pass validation), we calculate the estimated outer syndrome $\bfb^{(i)}$ of $\bfv$ with respect to $\mathsf{H}_\mathsf{c}$, and update the flag `found' to 1. If some $\bfv$ passes validation when `found' is set to 1 (meaning it is not the first to pass validation), we set the estimated outer syndrome to an erasure, i.e, $\bfb^{(i)} = \otimes$. Furthermore, if all decoding outputs $\bfv$ fail validation, an erasure  is also set to $\bfb^{(i)}$. This process, referred to as the \emph{inner decoding}, is repeated for every read $1\leq i \leq M$. Then, \emph{outer decoding} is performed by decoding the word of estimated outer syndromes $\underline{\bfb} \triangleq [\bfb^{(1)},\dots,\bfb^{(M)}]$, over $\mathsf{GF}(\nu)$, with respect to $\mathbf{H}_\mathsf{o}$ within the coset of syndrome $\bfS$, to form $\hat{\underline{\bfa}}$. Finally, the syndrome of each $\bfx_{\cI}^{(i)}$ with respect to the full-rank $\mathsf{H}$ is estimated by concatenating $\bfs^{(i)}$ sent by the encoder and $\hat{\bfa}^{(i)}$ resulted from outer decoding, and is then mapped to the \emph{single} word $\hat{\bfx}_{\cI}^{(i)}$ of this syndrome with respect to $\mathsf{H}$. The estimated read $\hat{\bfx}^{(i)}$ is finally reconstructed by inserting the read identifier $\bfw^{(i)}$ sent by the encoder to the suitable indices.
	
	The inner-decoder $\cD_1(\cdot,\cdot)$ finds the minimal-weight error word $\bfe$ for which $\mathsf{H}_1 \left(\bfy^{(i,j)}_{\cI} + \bfe\right)^T = \bfs_1^{(i)}$, and outputs $\bfy^{(i,j)}_{\cI}-\bfe$. This function can be easily implemented using standard syndrome-based decoders that do the same, only for the special case $\bfs_1^{(i)}=\mathbf{0}$. In this paper we assume that $\cD_1(\cdot,\cdot)$ is a \emph{bounded-distance decoder}~\cite{sps41}, discarding error words $\bfe$ with distance higher than half the code's minimum distance. Similarly, $\cD_{\mathsf{o}}(\bfb,\bfS)$, decodes $\bfb$ with respect to $\mathbf{H}_\mathsf{o}$ to a syndrome $\bfS$. $\cF_{\mathsf{H}}(\bfu)$ denotes the linear mapping of $\bfu$ to the word of syndrome $\bfu$ with respect to $\mathsf{H}$.
	
	We also note that the validation process includes cases of \emph{misvalidation}, that is, a vector $\bfv \neq \bfx^{(i)}_{\cI}$ passing validation, not to be confused with the case of a failed validation, happening when $\hat{\bfs}^{(i)}_2 \neq \bfs^{(i)}_2$, which forms a \emph{rejection} of the candidate.

	\begin{algorithm}[h!]
		\SetAlgoLined
		\textbf{Input}: $\cE\left(\{\bfx^{(i)}\}_{i=1}^M\right), \bfY, \mathsf{H}_1,\bar{\mathsf{H}}_2,\mathsf{H}_\mathsf{c}, \bfH_{\mathsf{o}}$ \\
		\For{$1\leq i \leq M$}{
			Align $\bfw^{(i)}$ over $\bfY$, and form $\mathrm{Y}^{(i)}$ (Eq. \ref{Eq:Candidates})\\
			Set $\text{'found'} \gets 0$ \\
			\For(\tcp*[h]{Inner Decoding}){$1 \leq j \leq |\mathrm{Y}^{(i)}|$}{
				Decode $\bfv = \cD_1(\bfy_{\cI}^{(i,j)},\bfs^{(i)}_1)$ \\
				Calculate $\hat{\bfs}^{(i)}_2 = \bar{\mathsf{H}}_2 \bfv^T$ \\
				\If(\tcp*[h]{Validation}){$\hat{\bfs}^{(i)}_2 = \bfs^{(i)}_2$}{
					\eIf{$\text{'found'}=0$}{
						Calculate $\bfb^{(i)} = \mathsf{H}_{\mathsf{c}}\bfv^T$, Set $\text{'found'} \gets 1$ \\ 
					}(\tcp*[h]{More Than One Candidate}){
						Set $\bfb^{(i)} = \bigotimes$, break
					}
				}
			}
			\If(\tcp*[h]{No Candidates}){$\text{'found'}=0$}{Set $\bfb^{(i)}=\bigotimes$}
		}
		\tcp*[h]{Outer Decoding} \\
		Decode $\hat{\underline{\bfa}} = \cD_{\mathsf{o}}(\underline{\bfb},\bfS)$, where $\underline{\bfb} = [\bfb^{(1)},\dots,\bfb^{(M)}]$\\
		\For(\tcp*[h]{Inverse Mapping}){$1 \leq i \leq M$}{
			Map $\hat{\bfx}_{\cI}^{(i)}=\cF_{\mathsf{H}}([\bfs^{(i)},\hat{\bfa}^{(i)}])$ \\
			Reconstruct $\hat{\bfx}^{(i)}$ from $\hat{\bfx}_{\cI}^{(i)} , \bfw^{(i)}$
		}
		\textbf{Output}: $\{\hat{\bfx}^{(i)}\}_{i=1}^{M}$		
		\caption{Construction \ref{cons:GenCons} Decoding}
		\label{Alg:Cons1Dec}
	\end{algorithm}	
\end{construction}
\vspace*{-5ex}
\begin{proposition}
	\label{prop:Rate}
	The rate of Construction~\ref{cons:GenCons} is given by
	\begin{equation*}
		\cR = 1-\frac{k_{\mathsf{o}}}{M}\cdot\frac{k_2-\ell}{n}.
	\end{equation*}
\end{proposition}

\begin{IEEEproof}
	Every $\bfw^{(i)},\bfs^{(i)}$ contain $\ell,(n-k_2)$ bits, respectively. $\bfS$ contains $M-k_{\mathsf{o}}$ elements, each represented by $\nu$ bits. Hence,
	\begin{align*}
		|\cE(\{\bfx^{(i)}\}_{i=1}^M)| & = M\cdot(\ell+n-k_2) + (M-k_{\mathsf{o}})\cdot \nu \\
		& = M\cdot n - k_{\mathsf{o}}\cdot (k_2-\ell) .
	\end{align*}
	Dividing by $M\cdot n$ information bits completes the proof.
\end{IEEEproof}

\begin{proposition}
	\label{prop:outer_success}
	Construction~\ref{cons:GenCons} yields an $(M,n,\cR,p,P_\mathsf{s})$-code if and only if at the outer-decoder output $\Pr\{\hat{\underline{\bfa}} = \underline{\bfa}\} \geq P_\mathsf{s}$. 
\end{proposition}

\begin{IEEEproof}
	By the definition of $\cF_{\mathsf{H}}(\cdot)$, we have $\hat{\bfx}_{\cI}^{(i)}= \bfx_{\cI}^{(i)}$ if and only if their syndromes with respect to $\mathsf{H}$ are identical. Furthermore, since $\bfw^{(i)}$ is known to the decoder, the former implies $\hat{\bfx}^{(i)} = \bfx^{(i)}$. Since $\bfs^{(i)}$ is also known to the decoder, if $\hat{\bfa}^{(i)} = \bfa^{(i)}$, the decoder has the correct syndrome bits with respect to $\mathsf{H}$, and can recover $\bfx^{(i)}$ perfectly.
\end{IEEEproof}

The design of the scheme's parameters for minimization of the compression rate is pursued in Section~\ref{Sec:SchemeAnalysis}.

\subsection{Employing Algebraic Codes in the Construction}
\label{SubSec:AlgCons}
Toward practically realizing the coding scheme, in this sub-section we study the use of known algebraic codes, with emphasis on adapting their use to source coding, and in particular to the requirements of Construction~\ref{cons:GenCons}.  Specifically, as the inner codes we will use the well-known binary primitive \emph{BCH codes}~\cite{sps27}. For the outer code, we will use the also very well-known \emph{Reed-Solomon (RS) codes}, which enjoy the desirable MDS (maximum distance separability) property. Both these code families have known efficient algebraic decoders (e.g., the Berlekamp-Massey algorithm~\cite{BerlekampE:84}), capable of correcting both errors and erasures~\cite{sps26}.  

The main challenge of using BCH codes is that their canonical representation is \emph{not} as binary parity-check matrices (as in Construction~\ref{cons:GenCons}), but rather given over a finite field $\mathsf{GF}(2^m)$, where $m$ is an integer that for code-length $n$ satisfies $n=2^m-1$. To guarantee the requirements of Construction~\ref{cons:GenCons}, we will focus on obtaining the following properties using BCH codes: (i) constructing a pair of codes $\{\cC_1,\cC_2\}$ whose syndromes are nested, (ii) implementing the inner decoder $\cD_1(\cdot,\cdot)$ using existing algebraic BCH decoders, and (iii) finding a complementary matrix $\mathsf{H}_{\mathsf{c}}$ ensuring $\mathsf{H}$ of full rank, preferably one offering efficient realization of the inverse mapping $\cF_{\mathsf{H}}$.

For $j\in\{1,2\}$, let $\cC_j = \mathsf{BCH}(n,k_j,2t_j+1)$ be a $[n,k_j]$ binary BCH code that can correct up to $t_j$ errors, such that $k_1 \geq k_2$. Property (iii) above can readily be obtained by a \emph{systematic} representation of the code $\cC_2$, a fact we show later in the section. However, a systematic parity-check matrix of a BCH code $\cC_2$ in general does not contain the rows of a systematic parity-check matrix for the BCH code $\cC_1$, which challenges Property (i). Furthermore, Property (ii) requires computation of algebraic syndromes (to be discussed later), which are different from the binary syndromes computed in Algorithm~\ref{Alg:Cons1Enc} and supplied to the decoder.

To solve this, we will show that one can choose arbitrary \emph{systematic} parity-check matrices for $\cC_1$ and $\cC_2$ in the encoder, and still be able to use standard syndrome-based algebraic BCH decoders. To do this, we exploit a type of nesting that BCH codes do have: with respect to their algebraic syndromes over $\mathsf{GF}(2^m)$. Let $\bfz = (z_0,z_1,\dots,z_{n-1})$ be a length-$n$ binary word. Its \emph{algebraic syndrome} is defined as follows~\cite{sps27}.

\begin{definition}
	\label{Def:AlgSyndrome}
	For every $x \in \mathsf{GF}(2^m)$, let $f_{\bfz}(x) \triangleq \sum_{i=0}^{n-1} z_i x^{i}$. Let $\alpha$ be a primitive element of $\mathsf{GF}(2^m)$. Then, the $t$-correcting \textbf{\emph{algebraic syndrome}} of $\bfz$ is defined by
	\begin{equation*}
		\algsynd(\bfz) = \left[f_{\bfz}(\alpha),f_{\bfz}(\alpha^2),\dots, f_{\bfz}(\alpha^{2t})\right] .
	\end{equation*}
\end{definition}
Note that the algebraic syndrome is a vector of length $2t$ over $\mathsf{GF}(2^m)$. For $j\in\{1,2\}$, denote by $\algsynd_j(\bfz)$ the algebraic syndrome of $\bfz$ with respect to $\cC_j$. Similarly, denote by $\binsynd_j(\bfz)$ the \emph{binary syndrome} of $\bfz$ with respect to $\cC_j$, computed by a systematic parity-check matrix $\mathsf{H}_j$. 
By the definition of the BCH construction, we have the following.
\begin{observation}\textbf{(Nested syndromes)}
	Let $\algsynd_1(\bfz),\algsynd_2(\bfz)$ be the algebraic syndromes of $\bfz$ with respect to $\cC_1,\cC_2$ respectively. Then, the first $2t_1$ elements of $\algsynd_2(\bfz)$ are equal to $\algsynd_1(\bfz)$.
\end{observation}

The following lemma shows that the algebraic syndrome of $\bfz$ can be recovered from its binary syndrome.
\begin{lemma}
	Let $\mathsf{H}_j$ be a parity-check matrix of $\cC_j$ having the identity matrix in its last $n-k_j$ columns. If $\binsynd_j = \mathsf{H}_j \bfz^T$, then it follows that $\algsynd_j(\bfz)=\algsynd_j(\textbf{0}\binsynd_j)$, where $\textbf{0}\binsynd_j \triangleq [0,\dots,0,\binsynd_j]$ of length $n$.
\end{lemma}
\begin{IEEEproof}
	Let us look at the following parity-check matrix
	\begin{align*}
		\mathsf{H}'_j = \begin{bmatrix} 
			1 & \alpha^1 & \alpha^2 & \dots & \alpha^n \\
			1 & (\alpha^2)^1 & (\alpha^2)^2 & \dots & (\alpha^2)^n \\
			\vdots & \vdots & \vdots & \ddots & \vdots \\
			1 & (\alpha^{2t_j})^1 & (\alpha^{2t_j})^2 & \dots & (\alpha^{2t_j})^n
		\end{bmatrix} ,
	\end{align*}
	where the $\mathsf{GF}(2^m)$ elements are taken as binary column vectors of length $m$. It is known that this matrix defines the binary BCH code $\cC_j$, albeit not in a systematic form~\cite{sps27}. Therefore, $\mathsf{H}_j$ and $\mathsf{H}'_j$ span the same linear space, implying there is a linear-transformation matrix $T_j$ such that $\mathsf{H}'_j = T_j \mathsf{H}_j$. Now, by the systematic form of $\mathsf{H}_j$, it is clear that $\mathsf{H}_j \bfz^T = \mathsf{H}_j (\textbf{0}\binsynd_j)^T = \binsynd_j$. Hence, we get
	\begin{align}
		\label{Eq:AlgSyndEqual}
		\mathsf{H}'_j \bfz^T = T_j \mathsf{H}_j \bfz^T = T_j \mathsf{H}_j (\textbf{0}\binsynd_j)^T = \mathsf{H}'_j (\textbf{0}\binsynd_j)^T .
	\end{align}
	Finally, it can be seen that the structure of $\mathsf{H}'_j$ is directly related to the algebraic syndrome, as we can write
	\begin{align*}
		\mathsf{H}'_j \begin{bmatrix} z_0 \\ z_1 \\ \vdots \\ z_{n-1} \end{bmatrix} = \sum_{i=0}^{n-1} z_i \begin{bmatrix} \alpha^i \\ (\alpha^2)^i \\ \vdots \\ (\alpha^{2t_j})^i \end{bmatrix} = \begin{bmatrix}
			\sum_{i=0}^{n-1} z_i (\alpha^{1})^i \\ \sum_{i=0}^{n-1} z_i (\alpha^2)^i \\ \vdots \\ \sum_{i=0}^{n-1} z_i (\alpha^{2t_j})^i \end{bmatrix} ,
	\end{align*}
	in which the $i$-th row exactly correspond to the $i$-th element of the algebraic syndrome, meaning $\mathsf{H}'_j \bfz^T = \algsynd_j(\bfz)$ (in binary representation). Combining with Equation~(\ref{Eq:AlgSyndEqual}), we get $\algsynd_j(\bfz) = \algsynd_j(\mathbf{0}\binsynd_j)$, completing the proof.
\end{IEEEproof}

Now, let $\cC_j=\mathsf{BCH}(n-\ell,k_j-\ell,2t_j+1)$ (possibly a shortened version of a primitive BCH code). When for $\cC_2$ we use a systematic parity-check matrix $\mathsf{H}_2$ (whose last $n-k_2$ columns form an identity matrix), for the complementary matrix we can use $\mathsf{H}_\mathsf{c} = [{I}_{k_2-\ell},\textit{0}]$, where ${I}_{k_2-\ell}$ denotes the identity matrix, and $\textit{0}$ denotes an all-zeros matrix of size $(k_2-\ell) \times (n-k_2)$. The vertical concatenation of $\mathsf{H}_2$ and $\mathsf{H}_\mathsf{c}$ is full rank, because both matrices are full rank, and given their structure no linear combination of rows of $\mathsf{H}_2$ can give a row of $\mathsf{H}_\mathsf{c}$ (and vice-versa). Furthermore, by using this matrix, the calculation of the syndrome $\bfa^{(i)}$ turns out to be simply extracting the first $k_2-\ell$ bits of $\bfx_{\cI}^{(i)}$.

\subsubsection*{Realizing Construction~\ref{cons:GenCons} with BCH and RS codes}
Based on the discussion above, we can now introduce the realization of the construction using algebraic codes.
Let $\cC_{\mathsf{o}} = \mathsf{RS}(M,k_{\mathsf{o}})$, where $\mathsf{RS}$ denotes a Reed-Solomon code. This code is taken over $\mathsf{GF}(2^{\mu})$, with $\log_2(M+1) \leq \mu \leq \nu$, by dividing the $\nu$ bits computed by $\mathsf{H}_\mathsf{c}$ into $\nu / \mu$ sections, and encoding each section separately into a $\cC_{\mathsf{o}}$ codeword. By keeping $\mu$ small enough (e.g. $\mu=8$), the field's order remains practical for processing regardless of the inner code's dimension.

The encoder remains almost identical to Algorithm~\ref{Alg:Cons1Enc}, with the following specifications:

\begin{itemize}
	\item $\bfs^{(i)}=\mathsf{H}_2 (\bfx_{\cI}^{(i)})^T$ (the binary syndrome) is computed with a systematic $\mathsf{H}_2$ for $\cC_2$,
	\item $\bfa^{(i)}$ is taken as the first $k_2-\ell$ bits of $\bfx_{\cI}^{(i)}$,
	\item $\bfS$ is computed as a concatenation of $\nu / \mu$ algebraic syndromes of $\cC_{\mathsf{o}}$.
\end{itemize}

The decoder is realized based on the following modifications to Algorithm~\ref{Alg:Cons1Dec}:
\begin{itemize}
	\item Calculate $\algsynd_2(\textbf{0}\bfs^{(i)})$ (the algebraic syndrome), and extract $\algsynd_1(\bfx_{\cI}^{(i)})$ as the first $2t_1$ $\mathsf{GF}(2^m)$ elements of it.
	\item Using an \emph{algebraic BCH decoder}, decode $\bar{\bfy}^{(i,j)}$  within the coset of the syndrome $\algsynd_1(\bfx_{\cI}^{(i)})$.
	\item Compute the \emph{algebraic} syndrome of the decoded word, $\algsynd_2(\bfv)$, and compare its last $2(t_2-t_1)$ elements to the corresponding elements in $\algsynd_2(\textbf{0}\bfs^{(i)})$ for validation.
	\item The mapping $\cF_{\mathsf{H}}([\bfs^{(i)},\hat{\bfa}^{(i)}])$ is realized by encoding $\hat{\bfa}^{(i)}$ to a codeword $\bfc\in\cC_2$ whose first $k_2-\ell$ bits equal $\hat{\bfa}^{(i)}$ (that can be easily done using $\mathsf{H}_2$), and then setting $\hat{\bfx}_{\cI}^{(i)} = \bfc + \textbf{0}\bfs^{(i)}$. The correctness of this step follows by the structures of $\mathsf{H}_2$,$\mathsf{H}_\mathsf{c}$ discussed above.
\end{itemize}



\section{A Distance for Alignment with Deletions}
\label{Sec:DelSemiMetric}

\begin{figure*}[h!]
	\centering
	\includegraphics[width=0.9\linewidth]{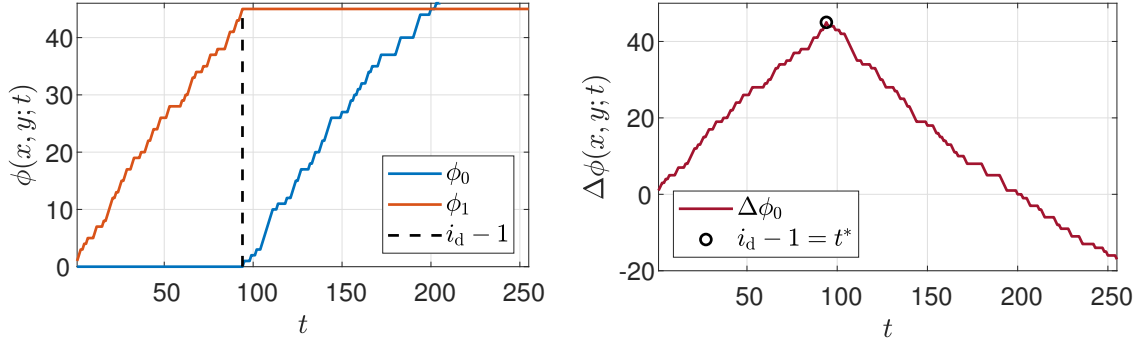}
	\caption{Illustration of the shift-compensating distance and its components, with no substitutions: cumulative distances (left), and their difference (right), with the deletion location minus 1 attaining the maximal difference.}
	\label{fig:OffsetDist}
\end{figure*}

To extend the coding scheme to also deal with deletion errors, we need to establish an efficient way to perform alignment (at the decoder) under such errors. With substitutions only, an alignment candidate is considered suitable if its Hamming distance to the read is small (see~\eqref{Eq:Candidates}). We now need an alternative distance measure that supports deletion errors between the reference and the sequenced reads. Ultimately, the distance measure proposed in this section will be used in the next section (Definition~\ref{Def:SCmatch}) to replace the alignment condition~\eqref{Eq:Candidates}. We focus on the case of a single deletion (and multiple substitutions) per inner block, and discuss extensions to multiple deletions (and to insertions) toward the end of the paper. Note that we can partition any read of given length into multiple inner blocks of any desired size, and therefore by keeping the inner-blocks short, a single deletion remains an interesting case in practice for a wide range of deletion probabilities.
\vspace*{-1.5ex}
\subsection{Existing Measures}
An immediate candidate for such a measure is the {\em Levenshtein distance}~\cite{sps28}, counting the minimal number of edits (deletions, insertions and substitutions) required to obtain one word from another. Nevertheless, this measure suffers from two main issues in our case: (i.) its complex calculation by dynamic-programming algorithms makes it impractical to evaluate each read along every possible offset in the reference, (ii.) allowing unrestricted error patterns, involving any combination of edits, introduces unfitting alignment candidates.


Another possible measure is the {\em shifted Hamming distance}~\cite{sps29}, which matches each read index with $r$ adjacent indices in the sub-sequence considered for alignment. For our purposes, since only deletions (and not insertions) are considered, it is sufficient to use shifts in one direction between the two input sub-sequences. This can be formalized by

\vspace*{-3ex}
\begin{align}\label{eq:shftH}
	\forall \bfx \in \bSig^n , \bfy \in \bSig^{n+r} : d_{\mathsf{SH}}(\bfx,\bfy) \triangleq \sum_{i=1}^n \bigwedge_{j=0}^r x_i \oplus y_{i+j} ,
\end{align}
where $\wedge$ denotes a logical `AND' operation, and $\bSig^m$ denotes a word of length $m$ from alphabet $\bSig$, and the binary operator $\oplus$ returns $0$ for equal symbols and $1$ otherwise. The main issue with this measure is that it allows low distance values from independent shifts between indices, ignoring the special shift structure of index deletion. This causes even random unrelated sequences to be declared close, increasing the number of improper alignments.

\subsection{Preliminaries}
The following definitions are the basic building blocks for our proposed distance measure.
\begin{definition} \textbf{(Cumulative Hamming distance)}
	\label{Def:CumHamDist}
	For $\bfx \in \bSig^n$, $\bfy \in \bSig^{n+r}$, define, for every $0\leq j \leq r$ and every $0\leq t\leq n$,
	\begin{align*}
		\phi_j (\bfx,\bfy ; t) \triangleq \sum_{i=1}^{t} x_i \oplus y_{i+j} ,
		\vspace*{-1.5ex}
	\end{align*}
	where $\oplus$ denotes an addition over $\mathsf{GF}(2)$. When clear from the context, we will denote $\phi_j (\bfx,\bfy ; t) = \phi_j(t)$.
\end{definition}

\begin{definition}
	Let $\bfx \in \bSig^n$, $\bfy \in \bSig^{n+r}$. Define, for every  $0\leq j \leq r-1$ and every $0 \leq t \leq n$,
	\begin{align*}
		\Delta \phi_j (\bfx,\bfy;t) & \triangleq \phi_{j+1}(\bfx,\bfy;t) - \phi_{j}(\bfx,\bfy;t) \\
		& = \sum_{i=1}^{t} [x_i \oplus y_{i+j+1}]-[x_i \oplus y_{i+j}] ,
	\end{align*}
	where by definition $\Delta \phi_j (\bfx,\bfy;0) = 0$. Again, when clear from the context, we will denote $\Delta\phi_j (\bfx,\bfy ; t) = \Delta\phi_j(t)$.
\end{definition}

\vspace*{-2.5ex}
\subsection{Single-Deletion Shift-Compensating Distance}
\vspace*{-0.5ex}
We can now introduce the desired distance measure. Intuitively, it measures the Hamming distance between two words, while compensating for a possible block shift caused by a single deletion occurring in one of them before entering the substitution channel.
\begin{definition}\label{def:shft_comp_measr}
	\label{Def:SCdist}
	Let $\bfx \in \bSig^n , \bfy \in \bSig^{n+1}$, and let
	\begin{align}
		\label{Eq:OptOCt}
		t^* \triangleq \argmax_{0\leq t \leq n} \{\Delta \phi_0 (\bfx,\bfy;t) \}.
	\end{align}
	Then, we define the \textbf{\emph{shift-compensating distance}} by
	\begin{align*}
		\sdeloc (\bfx, \bfy) & \triangleq \phi_1(\bfx,\bfy;n) - \phi_1(\bfx,\bfy;t^*) + \phi_0(\bfx,\bfy;t^*) \\
		& = \phi_1(\bfx,\bfy;n) - \max_{0\leq t \leq n} \{\Delta \phi_0 (\bfx,\bfy;t)\} \\
		& = \min_{0\leq t \leq n} \{\phi_1(\bfx,\bfy;n) - \Delta \phi_0 (\bfx,\bfy;t)\} .
	\end{align*}
\end{definition}

Note that $\sdeloc (\bfx, \bfy)\geq 0$, because $\phi_1(t)$ is non-decreasing and $\phi_0(t)$ is non-negative. Let us give some intuition to this measure through a simplified example. Let $\bfx$ be the word obtained from $\bfy$ by a deletion occurring in position $\isd$. Then, $x_i = y_i$ for every $1\leq i < \isd$, and $x_i = y_{i+1}$ for every $\isd \leq i \leq n$. Therefore $\phi_0(\isd-1) = 0$ and $\phi_1(n) - \phi_1(\isd-1) = 0$.
It follows that $\phi_1(n) - \Delta \phi_0(\isd-1) = 0$, and so it is clear that $(\isd-1) \in \argmax_{0\leq t \leq n} \{\Delta \phi_0(t) \}$ (otherwise implying a negative $\sdeloc (\bfx, \bfy)$). We conclude that: (i) an index maximizing $\Delta \phi_0$ gives an estimate for the deletion location (up to an ambiguity within a run of the same symbol around the deletion), and (ii) after estimating the deletion location, $\sdeloc$ gives the combined Hamming distance before and after this estimated location, while shifting the second part of $\bfx$ to compensate for the deletion. These conclusions are illustrated in Fig.~\ref{fig:OffsetDist}, where in the right plot it is evident that $t^*+1 = \isd$, thus locating the likely deletion position. The shift compensating measure sums the increments of $\phi_0$ (blue) left from $\isd-1$ and the increments of $\phi_1$ (red) right from $\isd$, which sum to zero since there are no substitutions. When introducing substitutions, these horizontal segments in $\phi_0,\phi_1$ will include random increases, with an average slope of $p$, attributed to the substitutions. 

\subsection{Similarity Properties}
In this sub-section we formalize the properties of $\sdeloc$. First, we examine its relation to the underlying Hamming distance.

\vspace*{-1.5ex}
\begin{lemma}
	\label{Lem:DistEquivalence}
	Let $\bfx \in \bSig^n , \bfy \in \bSig^{n+1}$. Let $\bfy^{[t]}$ denote the word obtained from $\bfy$ by a deletion in index $t$, for any $1\leq t \leq n+1$. Then,
	\vspace*{-3ex}
	\begin{align}\label{eq:delta_dist}
		\sdeloc (\bfx,\bfy) = \min_{1\leq t \leq n+1} \left\lbrace d_{\mathsf{H}} (\bfx,\bfy^{[t]}) \right\rbrace ,
		\\
		t^* = \argmin_{1\leq t \leq n+1} \left\lbrace d_{\mathsf{H}} (\bfx,\bfy^{[t]}) \right\rbrace - 1 .
	\end{align}
\end{lemma}

\begin{IEEEproof}
	It holds that
	\begin{align*}
		d_{\mathsf{H}}(\bfx,\bfy^{[t]}) & = \sum_{i=1}^{t-1} x_i \oplus y_i + \sum_{i=t}^{n} x_i \oplus y_{i+1} \\
		& = \phi_1(\bfx,\bfy;n) - \Delta\phi_0(\bfx,\bfy;t-1) .
	\end{align*}
	Minimizing over $t-1 \in \{0,\dots,n\}$ we get $\sdeloc(\bfx,\bfy)$ by definition, and the minimizing argument is exactly $t^*$.
\end{IEEEproof}

It is seen in~\eqref{eq:delta_dist} that $\sdeloc$ is the most natural distance for this error model: it gives the minimal Hamming distance between $\bfx$ and any single-deletion version of $\bfy$. Importantly, it has a linear calculation complexity making it more practical than other alternatives, such as the Levenshtein distance. 


Next, to analyze the new measure's alignment performance, in particular its ability to reject false alignment positions (thus reducing the count of improper alignments), we study its distribution when evaluated on a \emph{random} $\bfy$ word unrelated to $\bfx$. We first observe that for two independent random words $\bfx \in \bSig^{n}, \bfy\in \bSig^{n+1}$, we can write $\Delta \phi_0(\bfx,\bfy;t) = \sum_{i=0}^t \Delta_i , 1\leq t \leq n$, 
where $\Delta_0 = 0$, and $\{\Delta_i\}_{i=1}^t$ are independent random variables with support $\{-1,0,1\}$ and probabilities $\{0.25,0.5,0.25\}$, respectively. This type of sum forms a \emph{symmetric random walk with null steps}. Furthermore, we observe that in this case $\phi_1(\bfx,\bfy;n)$ is a Binomial random variable with $p=1/2$. Based on these observations, we get the following theorem.

\begin{theorem}
	\label{Th:RandomDistance}
	Let us denote by $\delta_n$ the random variable of $ \sdeloc(\bfx,\bfy)$ for randomly chosen $\bfx \in \bSig^{n},  \bfy \in \bSig^{n+1}$. Then,
	\begin{small}
		\begin{align*}
			P & (\delta_n = r) = \\
			& \frac{1}{4^n} \sum_{m=0}^{n-r} \sum_{t=m}^{n} \sum_{k=0}^{n-m} 
			\sum_{w = 0}^{k} \sum_{l =0}^{t-1} A_1(t,w,m,l) \smashoperator[lr]{\sum_{v=0}^{n-t-(k-w)}}A_2(v,t,k-w,r-l) ,
		\end{align*}
	\end{small}
	where we defined
	\begin{align*}
		A_1(t,w,m,l) \triangleq \frac{m}{t}\cdot {t \choose \alpha,\alpha+m,\beta, w - \beta} ,
	\end{align*}
	\begin{align*}
		A_2(v,t,k-w,r-l) = \frac{v+1}{\gamma + v + 1} {n-t \choose \gamma, \gamma + v,\eta, k-w-\eta} ,
	\end{align*}
	for $\alpha \triangleq (t-w-m)/2, \beta \triangleq l-\alpha, \gamma \triangleq (n-t-(k-w)-v)/2 , \eta \triangleq r-l-\gamma $, and ${z \choose u_1,u_2,\dots,u_m} \triangleq \frac{z!}{u_1!u_2!\dots u_m!}$, the multinomial coefficient.
\end{theorem}

\begin{IEEEproof}
	The full formal proof is given in Appendix~\ref{Ap:docRandom}. We include a sketch here. To count the number of sequence pairs that have $\sdeloc(\bfx,\bfy)=r$, we classify the pairs according to several variables appearing as summation indices. The primary variables are $t$ that is the value $t^*$ calculated in~\eqref{Eq:OptOCt}, and $m$ that is $\Delta \phi_0 (\bfx,\bfy;t^*)$. Given $t,m$, we count the number of random walks of $\Delta_i$ that attain a global maximum of $m$ at time $t$, and for each such walk expand the number of sequence pairs that have distance $r$. The variables $k$ and $w$ count the number of zeros of the random walk in total and until time $t$, respectively. The remaining summation indices are weight variables: $l$ for the sub-sequence $x_i \oplus y_i$ until time $t$, and $v$ for the sub-sequence $x_i \oplus y_{i+1}$ thereafter.
\end{IEEEproof}

\begin{figure}[h!]
	\centering
	\includegraphics[width=1\linewidth]{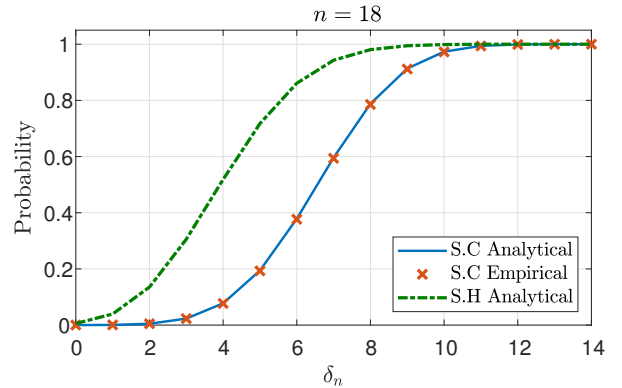}
	\caption{CDF of the proposed shift-compensating (S.C) distance (analytical and empirical) on unrelated sequences in comparison to the previously proposed shifted Hamming distance (S.H).}
	\label{Fig:SCrandSim}
\end{figure}

We plot in Fig.~\ref{Fig:SCrandSim} the distribution (CDF) of $\delta_n$ for $n=18$ obtained by Theorem~\ref{Th:RandomDistance} (solid). To validate the theoretical result, we include in the plot the empirical distribution collected from random sequences (red croses). We also show, in comparison, the CDF of the shifted-Hamming distance given in~\eqref{eq:shftH} (dashed green), demonstrating the significant advantage of $\sdeloc$ in obtaining high $\delta_n$ values for unrelated sequences.

\vspace*{-2ex}

\vspace*{-2ex}
\subsection{Metric-Like Properties}
In this section we show that our distance meets most of the conditions for being a metric. More specifically, we define a \emph{semi-metric}, which has all the properties of a standard metric~\cite{sps37}, short of a single technical property, and prove that our distance meets this definition. 

Let us denote $\bSig_a^b \triangleq \bigcup_{m=a}^b \bSig^m$, that is the set of all words with entries from $\bSig$ and lengths $m$ satisfying $a\leq m \leq b$. \vspace*{-1.5ex}
\begin{definition} \textbf{(Multi-length semi-metric)}
	\label{Def:SemiMetric}
	For every $0<a\leq b$, the function $\sd : \bSig_a^b \times \bSig_a^b \rightarrow [0,\infty)$ is called an $[a,b]$-\emph{length semi-metric} if it satisfies:
	\begin{enumerate}
		\item $\forall \bfx,\bfy \in \bSig_a^b : \sd(\bfx,\bfy) = \sd(\bfy,\bfx)$ (symmetry).
		\item $\forall \bfx,\bfy,\bfz \in \bSig_a^b $ such that $|\bfz|=|\bfx|$ or $|\bfz|=|\bfy|$, it follows that $ \sd(\bfx,\bfy) \leq \sd(\bfx,\bfz) + \sd(\bfz,\bfy)$ (triangle inequality).
		\item Let $\cY_0^m (\bfx) = \left\lbrace \bfy \in \bSig^m : \sd (\bfx,\bfy) = 0 \right\rbrace$. Then for every $\bfx \in \bSig^n$ we have $\cY_0^m(\bfx) = S_{|n-m|}(\bfx)$, as some deterministic set depending only on $|n-m|$ and the word $\bfx$, such that $S_0(\bfx) = \{\bfx\}$ (generalized identity of indiscernibles).
		
	\end{enumerate}
	These properties are similar to the properties defining a standard metric, with a generalization of the known identity of indiscernibles~\cite{sps38}, which will be useful as described after Proposition~\ref{Prop:SCisDist}.
\end{definition}

We now use $\sdeloc$ to form a natural symmetric distance, which is proved to be such a semi metric.

\begin{definition}
	Let $\bfx,\bfy \in \bSig_n^{n+1}$. The \textbf{\emph{symmetric shift-compensating distance}} is defined by
	\begin{align*}
		\sdoc(\bfx,\bfy) \triangleq \begin{cases}
			d_{\mathsf{H}}(\bfx,\bfy) & , |\bfx|=|\bfy| \\
			\sdeloc (\bfx, \bfy) & , |\bfx| = |\bfy| - 1 \\
			\sdeloc (\bfy, \bfx) & , |\bfx| = |\bfy| + 1
		\end{cases} .
	\end{align*}
\end{definition}

\begin{proposition}
	\label{Prop:SCisDist}
	The distance $\doc (\cdot,\cdot)$ is an $[n,n+1]$-length semi-metric for any $n>0$.
\end{proposition}

\begin{IEEEproof}
	We follow the conditions in Definition~\ref{Def:SemiMetric}. We denote by $\mathrm{D}_1(\bfx),\mathrm{I}_1(\bfx)$ the deletion and insertion balls of radius 1 of $\bfx$, respectively.
	\begin{enumerate}
		\item The symmetry $\doc(\bfx,\bfy) = \doc(\bfy,\bfx)$ is immediate from the definition.
		\item For the Hamming distance, the triangle inequality holds. For $|\bfy| = |\bfx|+1$ we examine two cases: (i.) Let $\bfz$ be some word such that $|\bfz|=|\bfx|$, then
		\begin{align*}
			\doc(\bfz,\bfy) = \min_{1\leq t\leq n+1} \{d_{\mathsf{H}}(\bfz,\bfy^{[t]} \} = d_{\mathsf{H}}(\bfz,\bfy^{[t_1]}),
		\end{align*}
		where $t_1$ is the minimizing index, and so
		\begin{align*}
			\doc(\bfx,\bfy) & \leq d_{\mathsf{H}}(\bfx,\bfy^{[t_1]}) \leq d_{\mathsf{H}}(\bfx,\bfz) + d_{\mathsf{H}}(\bfz,\bfy^{[t_1]}) \\
			& = \doc(\bfx,\bfz) + \doc(\bfz,\bfy) .
		\end{align*}
		(ii.) Let $\bfz$ be some word such that $|\bfz| = |\bfy|$. Similar to the former case $\doc(\bfx,\bfz) = d_{\mathsf{H}}(\bfx,\bfz^{[t_2]})$. Furthermore, for every $1\leq t \leq n+1$ we have $d_{\mathsf{H}}(\bfy^{[t]},\bfz^{[t]}) \leq d_{\mathsf{H}}(\bfy,\bfz)$, since an index is discarded in summation. Therefore,
		\begin{align*}
			\doc(\bfx,\bfy) & \leq d_{\mathsf{H}}(\bfx,\bfy^{[t_2]}) \leq d_{\mathsf{H}}(\bfx,\bfz^{[t_2]}) + d_{\mathsf{H}}(\bfz^{[t_2]},\bfy^{[t_2]}) \\
			& = \doc(\bfx,\bfz) + d_{\mathsf{H}}(\bfz^{[t_2]},\bfy^{[t_2]}) \\
			& \leq \doc(\bfx,\bfz) + \doc(\bfz,\bfy) .
		\end{align*}
		For $|\bfx|=|\bfy|=n$ and $|\bfz|=n+1$, 
		
		Hence, $\forall \bfx,\bfy,\bfz \in \bSig_n^{n+1}$, the triangle inequality holds.
		\item Based on Hamming distance properties we have $\cY_0^n(\bfx) = \{\bfx\}$. For some $\bfy \in \bSig^{n+1}$, based on Lemma~\ref{Lem:DistEquivalence}, we have $\sdeloc(\bfx,\bfy)=0$ if and only if $\exists t: d_\mathsf{H}(\bfx,\bfy^{[t]}) = 0$, i.e. $\bfx \in \mathrm{D}_1(\bfy)$. Since the latter occurs if and only if $\bfy \in \mathrm{I}_1(\bfx)$, we get $\cY_0^{n+1}(\bfx) = \mathrm{I}_1(\bfx) = \mathrm{I}_{|\bfy|-|\bfx|}(\bfx)$.
	\end{enumerate}
\end{IEEEproof}
The generalized identity of indiscernibles with $\cY_0^{n+1}(\bfx) = \mathrm{I}_1(\bfx)$ captures well the fact that multiple length-$(n+1)$ sequences are 1 deletion and 0 substitutions away from a length-$n$ alignment target $\bfx$, having $0$ distance as $\bfx$ itself.

\section{Extending The Scheme to Single-Deletion Multiple-Substitutions}
\label{Sec:SchemeDelExt}
\subsection{Channel Model}
In this section we extend the difference channel model to \emph{single-deletion multiple-substitutions (SDMS)} in each read. 

That is, for each read $\bfx^{(i)}$ there exists an index $r_i$ such that the read is obtained from $\bfy^{(i)} = Y_{r_i},\dots, Y_{r_i+n-1},Y_{r_i+n}$, of length $n+1$, by a single deletion and a certain number of substitutions. Equivalently, there is an integer $j_{\mathsf{d_i}}$ such that
\begin{align*}
	\bfx^{(i)} = \tilde{X}_{r_i}, \dots, \tilde{X}_{r_i+j_{\mathsf{d_i}}-2} , \tilde{X}_{r_i+j_{\mathsf{d_i}}},\dots,\tilde{X}_{r_i+n-1},\tilde{X}_{r_i+n} ,
\end{align*}

where $\tilde{\bfX}$ is the result of $\bfY$ passing through a substitution channel. In this model we have: (i) the $j_{\mathsf{d}}$-th index with respect to the segment, i.e. the $r_i + j_{\mathsf{d}} - 1$ with respect to the sequence, is deleted, and (ii) the last symbol in $\tilde{\bfX}^{(i)}$, i.e. $X_{r_i+n}$ is observed, keeping the length of the sequencer's output as $n$. This error model represents two equivalent scenarios: (i) the sequencer `skips' a symbol, while unknowingly keeping on sequencing an extra symbol to keep the length as defined, and (ii) the reference $\bfY$ includes an insertion with respect to the sequenced genome due to genomic diversity.
These result in the same relation between $\bfx^{(i)}$ and $\bfy^{(i)}$, and in both the \emph{sequencer is unaware whether a deletion occurred or not}.

\subsection{Alignment with Single-Deletion}
\label{SubSec:AlignMod}
A modification to the pre-decoding alignment described in Section~\ref{SubSec:Alignment} is required in order to deal with a potential deletion error in every read. This is done by simply replacing the Hamming distance with the proposed shift-compensating distance. To deal with the non-consecutive read identifier used for alignment, the calculation of $\phi_j(\bfx,\bfy)$ is easily extended to non-consecutive indices $1\leq i_1 < i_2 < \dots <i_{\ell}$ by
\begin{align*}
	\phi_j^{'}(\bfx,\bfy;t) = \sum_{k=1}^t x_{i_k} \oplus y_{i_k+j} \text{, \ for \ } 0 \leq t \leq \ell ,
\end{align*}
where consecutive index shifts (not restricted to the subset of indices sent to the decoder) are used only in $\bfy$, thus are available to the decoder. 
Denoting $\Delta \phi_j^{'}  \triangleq \phi_{j+1}^{'} - \phi_{j}^{'}$, $\sdeloct$ is now the corresponding modification of Definition~\ref{Def:SCdist}: 
\begin{equation}\label{eq:identifier_d}
	\sdeloct(\bfx,\bfy) \triangleq \min_{0\leq t \leq \ell} \{\phi_1^{'}(\bfx,\bfy;\ell) - \Delta \phi_0^{'} (\bfx,\bfy;t)\}.
\end{equation}

We now show an important property similar to Lemma~\ref{Lem:DistEquivalence}, showing that $\sdeloct$ captures the {\em read identifiers'} Hamming distance to the closest single-deletion vector.

\begin{lemma}
	\label{Lem:deloctUB}
	Let $\bfz\in\bSig^{n}$ be the word obtained from $\bfy \in \bSig^{n+1}$ by a deletion in index $\isd$, and let $\bfx\in\bSig^{n}$ be an arbitrary word. Then,
	\begin{align}
		\label{Eq:deloctUB}
		\sdeloct(\bfx,\bfy) \leq d_{\mathsf{H}} (f_\ell(\bfx),f_\ell(\bfz)).
	\end{align}
\end{lemma}
\begin{IEEEproof}
	We have
	\begin{align*}
		\phi_1^{'}(\bfx & ,\bfy ;  \ell) - \Delta\phi_{0}^{'}(\bfx,\bfy; t) = \\  \sum_{k=1}^{t} & x_{i_k} \oplus y_{i_k} + \sum_{k=t+1}^{\ell} x_{i_k} \oplus y_{i_k+1}
		= d_\mathsf{H} (f_\ell(\bfx), f_\ell(\bfy^{[i_{t+1}]})).
	\end{align*}
	Since $\sdeloct(\bfx,\bfy)$ is obtained by the minimization of this expression with respect to $t$, we get
	\begin{align*}
		\sdeloct(\bfx,\bfy) & = d_\mathsf{H} (f_\ell(\bfx), f_\ell(\bfy^{[i_{t^{*}+1}]})) \\ & \leq d_\mathsf{H} (f_\ell(\bfx), f_\ell(\bfy^{[i_{t'+1}]})) = d_\mathsf{H} (f_\ell(\bfx), f_\ell(\bfy^{[\isd]})) ,
	\end{align*}
	where $t'$ is the unique index such that $i_{t'}< \isd\leq  i_{t'+1}$. From this the statement follows.
\end{IEEEproof}

\begin{definition}
	\label{Def:SCmatch}
	Let $\bfx \in \bSig^n$, and let $\bfy \in \bSig^{n+1}$ be taken from some offset in $\bfY$. We say that $\bfy$ {\em \textbf{matches}} $\bfx$ if $\sdeloct (\bfx, \bfy)\leq \mathsf{T}$, for some integer distance threshold $\mathsf{T}$.
\end{definition} 

Now, by aligning the read identifier $\bfw^{(i)}$ to $\bfY$, each such \emph{match} yields a length-$(n+1)$ candidate $\bar{\bfy}^{(i,j)}$, and all matches form the set $\ \mathrm{Y}^{(i)}$. This view readily captures reads without deletions, which can be modeled as deleting the $(n+1)$-th index of $\bar{\bfy}^{(i,j)}$.

\subsection{Generating Sub-Candidates for Each Alignment Candidate}
\label{SubSec:SubCandMod}
In contrast with the substitutions-only case, we now need to account for a deletion in $\bar{\bfy}^{(i,j)}$ before decoding. Toward that, the extended scheme generates a list of sub-candidates $\{\bar{\bfy}^{(i,j)}_s\}$, according to the following procedure. Define $\chi$ to be a global integer tolerance parameter, and denote by $t\in\{0,\ldots,\ell\}$ the value of $t^{*}$ that minimized $\phi_1^{'}- \Delta\phi_{0}^{'}$ in~\eqref{eq:identifier_d} to obtain $\sdeloct(\bfx,\bar{\bfy}^{(i,j)})$. We denote here $[a,b]\triangleq \{a,a+1,\ldots,b\}$, and $(a,b]\triangleq \{a+1,\ldots,b\}$. Then the set $\{\bar{\bfy}^{(i,j)}_s\}$ is defined by all deletion indices $\xi$ that satisfy the following: 
\begin{align}
	\xi \in \cI_\chi(t) \triangleq \begin{cases}
		(i_{t-\chi}, i_{t+\chi+1}] & , \chi < t < \ell - \chi - 1 \\
		[1,i_{t+\chi+1}] & , 0\leq t \leq \chi \\
		(i_{t-\chi} , n+1] & , \ell - \chi - 1 \leq t \leq \ell
	\end{cases} ,
\end{align}
that is, all deletion-indices within $2\chi+1$ intervals of the read identifier's indices around $t$, with exceptions at the extremal indices. For $\chi=0 $, i.e., no tolerance, we have $\cI_0(t) = (i_t,i_{t+1}]$, whereas for $\chi \geq \max\{t,\ell-t\}$ we have $\cI_\chi(t) = [1,n+1]$, i.e., $\{\bar{\bfy}^{(i,j)}_s\}=\mathrm{D}_1(\bar{\bfy}^{(i,j)})$, where $\mathrm{D}_1(\cdot)$ denotes the single-deletion ball.
It is motivated to use $\chi>0$ because due to substitutions, the true deletion index in $\bar{\bfy}^{(i,j)}$ may fall outside its estimated interval defined by $t$. The value of $\chi$ controls the number of vectors qualifying to reconstruct $\bfx^{(i)}$ by inner decoding, and it may be set to best balance successful reconstruction from the proper alignments with effective rejection of false candidates. It can now be seen that by taking evenly spaced $\{i_1,\cdots,i_\ell \}$, undesirable large intervals can be avoided.

\subsection{Inner-code Decoding with Multiple Sub-Candidates}
\label{SubSec:InnerMod}
Recall from Section~\ref{subsec:coding_scheme} that in the substitutions-only case each candidate $\bar{\bfy}^{(i,j)}$ is passed to inner-code decoding and validation. Now with deletions, we need to decode and validate a {\em set of sub-candidates} $\{\bar{\bfy}^{(i,j)}_s\}$. To see how this should be done, we note the following observation. 
\vspace*{-1.5ex}
\begin{observation}
	\label{obs:ErrRegion}
	Let $\bfz^{[k]} \in \mathrm{D}_1(\bfz)$, and let $\bfx = \bfz^{[\isd]}$. Then, for $e_i \triangleq \bfz_i \oplus \bfz_{i+1}$, \vspace*{-1.5ex}
	\begin{align*}
		\bfx_i \oplus \bfz_i^{[k]} = \begin{cases}
			0 & , i<\min(k,\isd) \text{ or } i\geq\max(k,\isd) \\
			e_i & , \text{otherwise}
		\end{cases}.
	\end{align*}
\end{observation}

\begin{figure}[h!]
	\centering
	\includegraphics[width=0.65\linewidth]{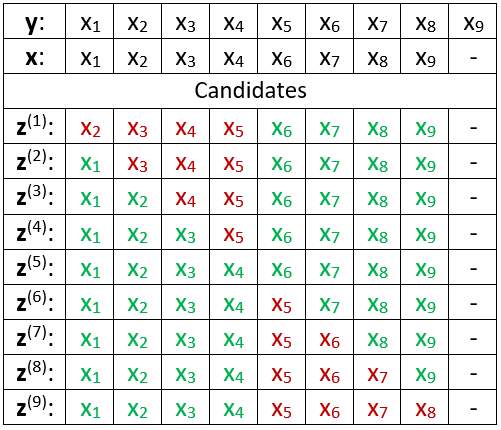}
	\caption{Illustration of the error regions (in red) for different vectors within a deletion ball, where $\bfx = \bfz^{(5)}$. }
	\vspace*{-2ex}
	\label{Fig:DelBallDec}
\end{figure}

This means that if $\isd$ is the actual deletion index and $k$ is the one chosen for some sub-candidate, this mismatch may introduce errors only in the region between those two indices, defined as the \emph{error region}, as illustrated in Fig.~\ref{Fig:DelBallDec}. 

It is thus likely that for the proper alignment of $\bfx^{(i)}$, multiple sub-candidates with deletion indices close to $\isd$ will correctly decode to $\bfx^{(i)}$, even if $\bfx^{(i)}$ contains some substitutions with respect to $\bfy^{[\isd]}$. This motivates the following treatment of $\{\bar{\bfy}^{(i,j)}_s\}$ in the modified decoder, specified in Algorithm~\ref{Alg:Cons1ModDec}. For every candidate $\bar{\bfy}^{(i,j)} \in \mathrm{Y}^{(i)}$: (i.) Decode and validate every sub-candidate $\bfu\in \{\bar{\bfy}^{(i,j)}_s\}$, (ii.) store any $\bfv$ that was successfully decoded and validated, and the number of its appearances $A(\bfv)$ along the decoding instances in $\{\bar{\bfy}^{(i,j)}_s\}$, (iii.) apply a majority rule
\begin{align}
	\label{Eq:Majority}
	\mathsf{Maj}(\cV) = \begin{cases}
		\bfv^* & , \forall \bfv \in \cV \setminus \{\bfv^* \}:  A(\bfv^*) > A(\bfv) \\
		\emptyset & , \text{there exist no such } \bfv^*
	\end{cases} .
\end{align}
From this point, $\bfv^*$ takes the role of $\bfv$ as defined in the decoding of Construction~\ref{cons:GenCons}, and the rest of the decoding process is unaltered. The encoding process is also unchanged. 

\begin{algorithm}[h!]
	\SetAlgoLined
	\textbf{Input}: $\cE\left(\{\bfx^{(i)}\}_{i=1}^M\right), \bfY, \mathsf{H}_1,\bar{\mathsf{H}}_2,\mathsf{H}_\mathsf{c}, \bfH_{\mathsf{o}}$ \\
	\For{$1\leq i \leq M$}{
		Align $\bfw^{(i)}$ over $\bfY$, and form $\mathrm{Y}^{(i)}$\\
		Set $\text{'found'} \gets 0$ \\
		\For(\tcp*[h]{Inner Decoding}){$1 \leq j \leq |\mathrm{Y}^{(i)}|$}{
			Set $\cV = \emptyset$ \\
			\For{every $\bfu \in \{\bar{\bfy}^{(i,j)}_s\}$}{
				Decode $\bfv = \cD_1(\bfu_{\cI},\bfs^{(i)}_1)$ \\
				Calculate $\hat{\bfs}^{(i)}_2 = \bar{\mathsf{H}}_2 \bfv^T$	\\
				\If(\tcp*[h]{Validation}){$\hat{\bfs}^{(i)}_2 = \bfs^{(i)}_2$}{
					$\cV \gets \cV \cup \{\bfv\}$\\
					$A(\bfv) = A(\bfv) + 1$
				}
			}
			$\bfv^* = \mathsf{Maj}(\cV)$ (Eq.~\ref{Eq:Majority}) \\
			\If(\tcp*[h]{Appropriate candidate}){$\bfv^* \neq \emptyset$}{
				\eIf{$\text{'found'}=0$}{
					Calculate $\bfb^{(i)} = \mathsf{H}_{\mathsf{c}}(\bfv^*)^T$\\
					Set $\text{'found'} \gets 1$ \\ 
				}(\tcp*[h]{More Than One Candidate}){
					Set $\bfb^{(i)} = \bigotimes$, break
				}
			}
		}
		\If(\tcp*[h]{No Candidates}){$\text{'found'}=0$}{Set $\bfb^{(i)}=\bigotimes$}
	}
	\tcp*[h]{Outer Decoding} \\
	Decode $\hat{\underline{\bfa}} = \cD_{\mathsf{o}}(\underline{\bfb},\bfS)$, where $\underline{\bfb} = [\bfb^{(1)},\dots,\bfb^{(M)}]$\\
	\For(\tcp*[h]{Inverse Mapping}){$1 \leq i \leq M$}{
		Map $\hat{\bfx}_{\cI}^{(i)}=\cF_{\mathsf{H}}([\bfs^{(i)},\hat{\bfa}^{(i)}])$ \\
		Reconstruct $\hat{\bfx}^{(i)}$ from $\hat{\bfx}_{\cI}^{(i)} , \bfw^{(i)}$
	}
	\textbf{Output}: $\{\hat{\bfx}^{(i)}\}_{i=1}^{M}$		
	\caption{Decoding Construction \ref{cons:GenCons} with Deletions}
	\label{Alg:Cons1ModDec}
\end{algorithm}

	%

\section{Probabilistic Analysis for Setting Code Parameters}
\label{Sec:SchemeAnalysis}
In this section we analyze the compression scheme proposed in Section~\ref{Sec:CodeConst} when used with $t$-correcting inner codes and MDS outer codes. The main purpose of this analysis is to serve tools that can help in setting the code and alignment parameters. Since the compression rate of the scheme is fixed by the code parameters, we focus on analyzing the probability that the block of reads is successfully reconstructed by the decoder with no error. Throughout this section, we model the genome as an i.i.d. Bernoulli$(1/2)$ sequence, and in addition assume the relation $\bfx^{(i)} = \mathsf{S}_{2,n}(\bfy^{(i)},p)$ (Lemma~\ref{Lem:EquivChan}) for the proper alignment.
Extension of the analysis to other genome statistical models can be done based on related studies such as \cite{sps5,sps7}. 

We begin our anaylsis for the case of substitution-errors only, and extend it to the case of SDMS errors in Sections~\ref{subsec:ModAlignmentAnalysis} and~\ref{SubSec:ModInnerAnalysis}. The analysis has two main components: first analyzing the \emph{inner-code performance} as a function of the alignment parameters and the inner-code parameters, and then the \emph{outer-code performance} as a function of the inner-code performance and the outer-code parameters. The outer-code analysis is the same for both error models.

For the analysis, we denote by $\mathsf{f}_{\mathsf{b}}(n,p,t) \triangleq {n \choose t} p^t (1-p)^{n-t} $, $\mathsf{F}_{\mathsf{b}}(n,p,t) \triangleq \sum_{i=0}^{t} \mathsf{f}_{\mathsf{b}}(n,p,i)$, the \textit{probability mass function} (\textit{PMF}) and \textit{cumulative distribution function} (\textit{CDF}) of a binomial random variable with parameters $(n,p)$, evaluated at the value of $t$, respectively. We also denote $V_n(t) \triangleq \sum_{i=0}^{t} {n \choose i}$ as the volume of a Hamming ball with radius $t$ of length-$n$ words. Finally, we denote $n_\ell \triangleq n-\ell$.
\vspace*{-2ex}
\subsection{Inner-Code Analysis}
\label{subsec:InnerAnalysis}
The inner decoder is invoked in Algorithm~\ref{Alg:Cons1Dec} on both the proper alignment of $\bfx^{(i)}$ in $\bfY$ (if found), denoted by $\bfy^{(i)}$, and, possibly, on improperly aligned $\bar{\bfy}^{(i,j)}$ vectors that are not related to $\bfx^{(i)}$. The scheme's performance depends on the inner-decoding outcomes for both types of inputs. We begin with the proper alignment.
\vspace*{-2ex}
\begin{Definition}
	\label{Def:InnerProperProb}
	The following probabilities are defined:
	\begin{itemize}
		\item $\Ppnexc$ - for succeeding in both alignment and decoding, i.e., $\bfy^{(i)} \in \mathrm{Y}^{(i)}$ and $\bfv = \bfx^{(i)}_{\cI}$,
		\item $\Ppmiv$ - for succeeding in alignment and misvalidating the reconstruction, i.e., $\bfy^{(i)} \in \mathrm{Y}^{(i)}$, $\bfv \neq \bfx^{(i)}_{\cI}$ and $\hat{\bfs}_2^{(i)} = \bfs_2^{(i)}$,
		\item $\Ppfail$ - for failing in either alignment or validation, i.e., $\bfy^{(i)} \notin \mathrm{Y}^{(i)}$ or $\hat{\bfs}_2^{(i)} \neq \bfs_2^{(i)}$.
	\end{itemize}
	
\end{Definition}
As seen above, we use the superscript $\mathsf{(p)}$ to represent probabilities pertaining to the \emph{proper} alignment. The probabilities in Definition~\ref{Def:InnerProperProb} describe every possible outcome of the proper-alignment inner decoding, hence $\Ppfail = 1-\Ppnexc-\Ppmiv$. Therefore, it suffices to evaluate $\Ppnexc$ and $\Ppmiv$.

\begin{lemma}\label{Lem:Sucp}
	Let $t_1 \triangleq \left\lfloor (d_1-1) / 2 \right\rfloor$. Then we have
	\begin{align*}
		\Ppnexc = \mathsf{F_b}(\ell,p,\mathsf{T}) \cdot \mathsf{F_b}(n_\ell,p, t_1).
	\end{align*}
\end{lemma}

\begin{IEEEproof}
	The first term is the probability of not exceeding the threshold of $\mathsf{T}$ substitutions in the read identifier. The second term is the probability of not exceeding the substitution-correction capability $t_1$ of $\cC_1$ in the inner decoding. Since the indices of alignment and inner decoding are disjoint, the events $\bfy^{(i)} \in \mathrm{Y}^{(i)}$ and $\bfv = \bfx^{(i)}_{\cI}$ are independent, and we get a multiplicative joint probability. 
\end{IEEEproof}

For the calculation of $\Ppmiv$, we detail the properties of decoding and validating within specific code cosets.

\begin{definition}
	\label{Def:Coset}
	Let $\cC$ be a linear code with a parity-check matrix $\mathsf{A}$. Then, let $\cC^{[\bfs]}$ be the coset of the code $\cC$ with syndrome $\bfs$, given formally by
	\begin{align*}
		\cC^{[\bfs]} = \left\lbrace \bfv = \bfc + \bfu_\bfs \ | \ \bfc \in \cC \right\rbrace ,
	\end{align*}
	where $\bfu_\bfs$ is a minimal-weight word $\bfu$ such that $\mathsf{A} \bfu^T = \bfs$.
\end{definition}

We use the following well-known lemma, which we include with proof due to our non-standard notation.
\begin{lemma}
	\label{Lem:CosetsSubset}
	Let $\bfs = [\bfs_1,\bfs_2]$ be a syndrome of $\cC_2$ with $\bfs_1, \bfs_2$ corresponding to $\mathsf{H}_1, \bar{\mathsf{H}}_2$, respectively. Then $\cC_2^{[\bfs]} \subseteq \cC_1^{[\bfs_1]}$.
\end{lemma}

\begin{IEEEproof}
	By the nested structure it is known that $\cC_2 \subseteq \cC_1$, and for every $\bfv$ such that $\mathsf{H}_2 \bfv^T = \bfs$, we get $\mathsf{H}_1 \bfv^T = \bfs_1$. Since $\mathsf{H}_2 \bfu_\bfs^T = \bfs$, we get $\mathsf{H}_1 \bfu_\bfs^T = \bfs_1$, therefore $\bfu_\bfs \in \cC_1^{[\bfs_1]}$ and can be written as $\bfu_\bfs = \bfc + \bfu_{\bfs_1}$, where $\bfc \in \cC_1$. Hence, for any word $\bfv \in \cC_2^{[\bfs]}$, we have that $\bfv = \bfc' + \bfu_\bfs = \bfc' + \bfc + \bfu_{\bfs_1} = \bfc'' + \bfu_{\bfs_1}$, where $\bfc',\bfc'' \in \cC_1$. By the latter, we get $\bfv \in \cC_1^{[\bfs_1]}$.
\end{IEEEproof}

We now analyze the probability of misvalidating a random word, which will be used in an upper bound for $\Ppmiv$.

\begin{Definition}
	\label{Def:MivProb}
	Let $\bfv \in \cC_2^{[\bfs]} \subseteq \cC_1^{[\bfs_1]}$ with $\bfs = [\bfs_1,\bfs_2]$, and let $\bfy_{\cI}$ be a random word chosen uniformly from the entire space $\{0,1\}^{n_{\ell}}$. Then the {\em probability of misvalidating} $\bfy_{\cI}$, denoted $\Prmiv$, is the probability of miscorrecting $\bfy_{\cI}$ with the decoder of $\cC_1$ to some $\bfv'\neq \bfv$, such that $\bar{\mathsf{H}}_2 (\bfv')^T = \bar{\mathsf{H}}_2 \bfv^T = \bfs_2$.
\end{Definition}

\begin{lemma}\label{Lem:MivRand}
	The probability of misvalidating a random word satisfies
	\begin{align}\label{Eq:randMiv}
		\Prmiv = \frac{(2^{k_2-\ell}-1)\cdot V_{n_\ell}(t_1)}{2^{n_\ell}} .
	\end{align}
\end{lemma}

\begin{IEEEproof}
	Let $\bfv \in \cC_2^{[\bfs]}$, and let $\bfy_{\cI}$ be a random word. Then, by Definition~\ref{Def:MivProb}, the misvalidation probability is the probability of $\bfy_{\cI}$ being decoded (miscorrected) by the \emph{bounded-distance decoder} of $\cC_1$ to a word within $\cC_2^{[\bfs]} \setminus \{\bfv\}$. Formally, define the set
	\begin{align}
		\label{Eq:C1MivCandidates}
		\cC_2^{[\bfs]}(t_1,\bfv) \triangleq \left\lbrace \bfg \ | \ \exists \bfv' \in \cC_2^{[\bfs]} \setminus \{\bfv\} : d_{\mathsf{H}}(\bfg,\bfv') \leq t_1 \right\rbrace,
	\end{align}
	for which the condition for misvalidation is: $\bfy_{\cI}\in \cC_2^{[\bfs]}(t_1,\bfv)$.   	 
	The size of this set is $\left| \cC_2^{[\bfs]}(t_1,\bfv) \right| = (|\cC_2|-1)\cdot V_{n_\ell}(t_1) = (2^{k_2-\ell}-1)\cdot V_{n_\ell}(t_1)$, because the balls of radius $t_1$ do not intersect,
	and the misvalidation probability is thus
	\begin{align*}
		\Prmiv = \frac{ \left| \cC_2^{[\bfs]}(t_1,\bfv) \right|}{2^{n_\ell}} = \frac{(2^{k_2-\ell}-1)\cdot V_{n_\ell}(t_1)}{2^{n_\ell}} .
	\end{align*}
\end{IEEEproof}

For the following, we assume that $\cC_1$ is a \emph{proper code}~\cite{sps16}, a definition that covers any code whose probability of undetected error (miscorrection) is monotonically increasing with the substitution probability of the channel. This restriction is only for simplifying the probability expressions, and it is in no way necessary for practically reaching performance targets using the subsequent analysis. Moreover, many codes are proper for most of the distance range (e.g. \textit{BCH}~\cite{sps15}, \textit{LDPC}~\cite{sps44}). We can now complete the analysis of the proper alignment.
\begin{lemma}
	\label{Lem:ProperMiv}
	The probability of misvalidating the proper alignment satisfies
	\begin{align}
		\Ppmiv \leq \mathsf{F_b}(\ell,p,\mathsf{T}) \cdot \Prmiv .
	\end{align}
\end{lemma}

\begin{IEEEproof}
	The first term is again the probability of successful alignment. Due to the properness of $\cC_1$, for every $p \leq 1/2$, the probability of miscorrecting the noisy read, i.e., having it fall within a Hamming ball of radius $t_1$ centered on some codeword of $\cC_1$, is upper bounded by the probability of miscorrecting a random word.
\end{IEEEproof}

Now, we move to analyze improper alignments, and our main objective is upper bounding the probability that one or more improper alignments will be misvalidated, an event that induces either an erasure or an error in the outer codeword. 

We denote by $\Pmgz$ the probability of having at least one misvalidated improper alignment (accepted at alignment as candidate, and then miscorrected and validated).


\begin{lemma}
	\label{Lem:MivGreaterThanZero}
	It holds that
	\begin{align*}
		\Pmgz \leq (L-n) \cdot \mathsf{F_b}(\ell,1/2,\mathsf{T}) \cdot \Prmiv + (n-1) \cdot \Prmiv.
	\end{align*} 
\end{lemma}
\vspace*{-2ex}
\begin{IEEEproof}
	Let $\bfx$ denote a read corresponding to some starting index $r'$, and aligned over the reference $\bfY$. Let $\mathbb{I}_r$ denote the indicator of the event in which $Y_r,\dots,Y_{r+n-1}$ is improperly aligned, and misvalidated with respect to $\bfx$, for any $r \in \cJ \triangleq \{1,\dots,L\}\setminus \{ r' \} $. Then,
	\begin{align*}
		\Pmgz = \Pr \left\lbrace \sum_{r \in \cJ} \mathbb{I}_r > 0 \right\rbrace \leq \sum_{r \in \cJ} \Pr \left\lbrace \mathbb{I}_r > 0 \right\rbrace ,
	\end{align*}
	where the inequality is due to the union bound \cite{sps46}. Note that the probability $\Pr \left\lbrace \mathbb{I}_r > 0 \right\rbrace$ is conditioned on the fact that $\bfx= \mathsf{S}_{2,n}(Y_{r'},\dots,Y_{r'+n-1},p)$, for some $r'\neq r$ (the proper alignment). We keep this conditioning implicit, but take it into account in the derivation of $\Pr \left\lbrace \mathbb{I}_r > 0 \right\rbrace$. Assume without loss of generality that $r'>r$ (the other direction follows similarly from symmetry), and denote $\alioff=r'-r$. If $\alioff\geq n$ (no overlap), the conditioning can be ignored because on disjoint indices $\bfx$ and $Y_r,\dots,Y_{r+n-1}$ are independent Bernoulli$(1/2)$ vectors. Thus we get 
	\begin{align}
		\label{Eq:NonOverlapMivProb}
		\forall r \ | \ \alioff\geq n :  \Pr \left\lbrace \mathbb{I}_r > 0 \right\rbrace = \mathsf{F_b}(\ell,1/2,\mathsf{T}) \cdot \Prmiv,
	\end{align}
	where the independence between alignment and misvalidation follows from their disjoint indices.
	In the case $\alioff< n$ we need to be more careful, because due to overlap the vectors $\bfx$ and $Y_r,\dots,Y_{r+n-1}$ are no longer independent. Denote $\bfy\triangleq [Y_r,\dots,Y_{r+n-1}]$, $\bfy'\triangleq [Y_{r'},\dots,Y_{r'+n-1}]$. We now prove that $\bfy-\bfy'$ is uniformly distributed on the set $\{0,1\}^{n}$ when the sequence $Y_r,\ldots,Y_{r'},\ldots,Y_{r+n-1},\ldots,Y_{r'+n-1}$ is drawn i.i.d. Bernoulli$(1/2)$, as assumed in the model. To do this, we fix an arbitrary $\dify\in\{0,1\}^{n}$, and calculate the probability that $\bfy-\bfy'=\dify$, where subtraction is bitwise modulo $2$.
	\begin{align*}
		\Pr & \left\lbrace \bfy-\bfy' =\dify\right\rbrace = \Pr \left\lbrace \right.\\
		& Y_{r+\alioff} = Y_{r} - z_1 , Y_{r+1+\alioff}=Y_{r+1} - z_2 ,\ldots,\\
		& \ \ Y_{r+2\alioff-1}=Y_{r+\alioff-1} - z_{\alioff}, Y_{r+2\alioff}=Y_{r+\alioff} - z_{\alioff+1},\ldots, \\
		& \ \ \left. Y_{r+n-1+\alioff}=Y_{r+n-1} - z_{n}\right\rbrace.
	\end{align*}
	From the uniform distribution of $Y_r,\ldots,Y_{r'},\ldots,Y_{r+n-1},\ldots,Y_{r'+n-1}$, we can rewrite the same probability as
	\begin{align*}
		\Pr & \left\lbrace \bfy-\bfy'=\dify\right\rbrace = \frac{1}{2^{n+{\alioff}}}\sum_{Y_{r},\ldots,Y_{r+\alioff+n-1}} \mathbb{I} \left\lbrace \right.\\
		& Y_{r+\alioff} = Y_{r} - z_1 , Y_{r+1+\alioff}=Y_{r+1} - z_2,\ldots,\\
		& \ \ Y_{r+2\alioff-1}=Y_{r+\alioff-1} - z_{\alioff}, Y_{r+2\alioff}=Y_{r+\alioff} - z_{\alioff+1},\ldots, \\
		& \ \ \left. Y_{r+n-1+\alioff}=Y_{r+n-1} - z_{n}\right\rbrace.
	\end{align*}
	It can be observed that when fixing the \emph{sub}sequence $Y_{r}=y_1,Y_{r+1}=y_2,\ldots,Y_{r+\alioff-1}=y_{\alioff}$, the indicator function is evaluated to $1$ for exactly one sequence that satisfies all the constraints in the argument (left-hand side values up to $Y_{r+2\alioff-1}$ are set by $y_1,\ldots,y_{\alioff}$ and $z_1,\ldots,z_{\alioff}$, and from $Y_{r+2\alioff}$ onward they are set by the previously set values and $z_{\alioff+1},\ldots,z_{n}$). When summing over all assignments to $y_1,\ldots,y_{\alioff}$, we get 
	\begin{align*}
		\Pr \left\lbrace \bfy-\bfy'=\dify\right\rbrace = \frac{2^{\alioff}}{2^{n+{\alioff}}} = \frac{1}{2^{n}}, 
	\end{align*}
	proving the uniform distribution of $\dify$. 
	
	The uniform distribution of the difference between $\bfy$ and $\bfy'$ implies that the validation outcome of $\bfy$ is independent from the fact that $\bfx$ (with syndrome $\bfs$) is likely proximate to $\bfy'$. We can thus invoke Lemma~\ref{Lem:MivRand}, and get 
	
	\begin{align}
		\label{Eq:OverlapMivProb}
		\forall r \ | \ \alioff<n : \Pr \left\lbrace \mathbb{I}_r > 0 \right\rbrace \leq \Prmiv , 
	\end{align}
	where in this case we upper bound the alignment-success probability by $1$, due to possible dependence between overlapping identifier bits when conditioning on the proper alignment. Since there are $n-1$ indices with $\alioff < n$, and $L-n$ indices for which $\alioff \geq n$, we get the desired upper bound. 
\end{IEEEproof}


\subsection{Outer-Code Analysis}
\label{SubSec:OuterChan}
In the outer code, the syndromes with respect to $\mathsf{H}_{\mathsf{c}}$ are treated as symbols in $\mathsf{GF}(2^\nu)$, which may be erased depending on the outcome of the inner decoding. Furthermore, a misvalidated inner-decoder output $\bfv$ may introduce an erroneous symbol. The outer-decoded word $\underline{\bfb}$ can thus be modeled as transmitting $\underline{\bfa}$ through an \emph{outer channel} introducing erasures and errors. The probabilities of these events are directly induced by the probabilities of the inner-decoder outcomes, analyzed in the previous sub-section. We can now examine the erasure and error probabilities of the outer channel, denoted by $\Pers,\Perr$, respectively.  

\begin{lemma}
\label{Lem:OuterProbUB}
The following upper bounds hold
\begin{align*}
	\Pers & \leq (1-\Ppfail) \cdot \Pmgz + \Ppfail , \\
	\Perr & \leq \Ppfail \cdot \Pmgz + \Ppmiv.
\end{align*}
\end{lemma}

\begin{IEEEproof}
An erasure occurs if there are no validated candidates (i.e., the proper alignment fails and no improper alignment is misvalidated), or more than one candidate has been validated. Hence,
\begin{align*}
	\Pers & = (1-\Ppfail) \Pmgz + \Ppfail (1-\Pmgz) \\
	& \leq (1-\Ppfail) \cdot \Pmgz + \Ppfail .
\end{align*}

An error occurs if the proper alignment is misvalidated and no improper alignment is misvalidated, or the proper alignment fails and exactly one improper alignment is misvalidated. The probability of exactly one improper misvalidation is upper bounded by $\Pmgz$. Hence,
\begin{align*}
	\Perr & \leq \Ppfail \cdot \Pmgz + \Ppmiv \cdot (1-\Pmgz) \\ 
	& \leq \Ppfail \cdot \Pmgz + \Ppmiv.
\end{align*}
\end{IEEEproof}

\begin{figure*}[h!]
\centering
\begin{subfigure}{.45\textwidth}
	\centering
	\includegraphics[width=\linewidth]{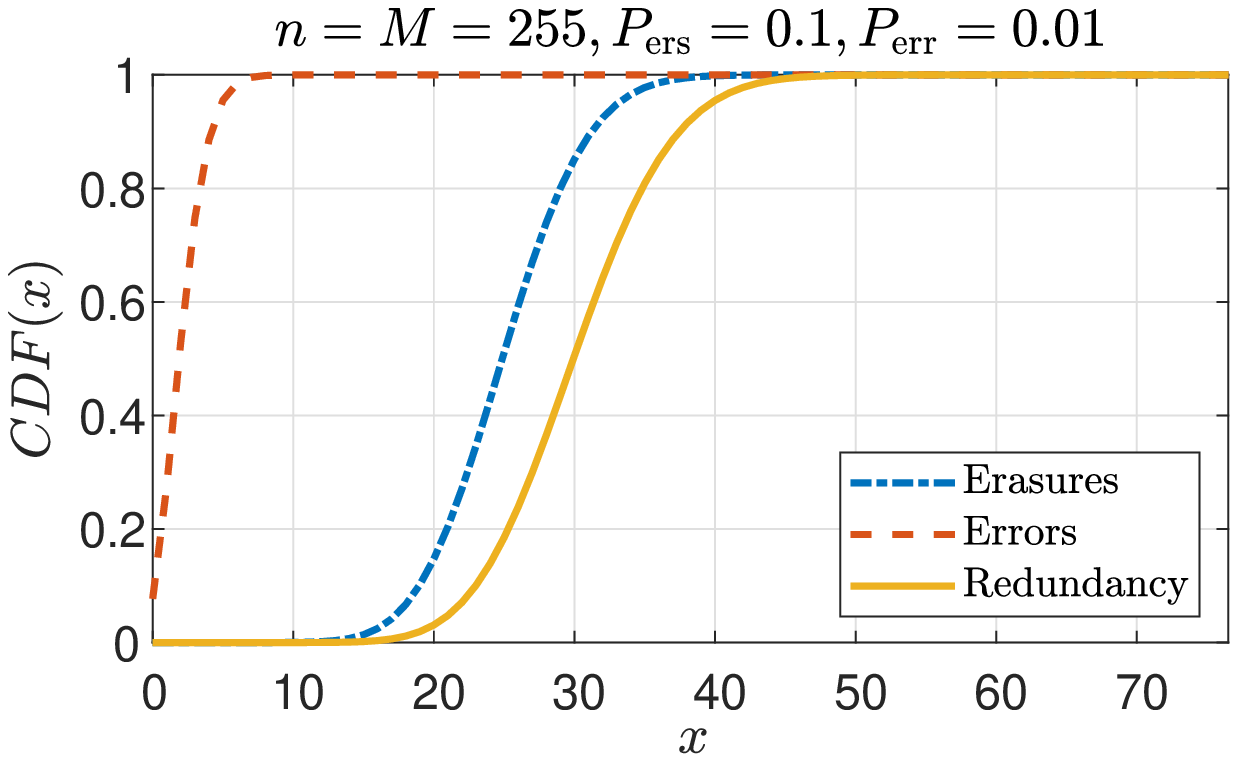}
\end{subfigure}
\begin{subfigure}{.45\textwidth}
	\centering
	\includegraphics[width=\linewidth]{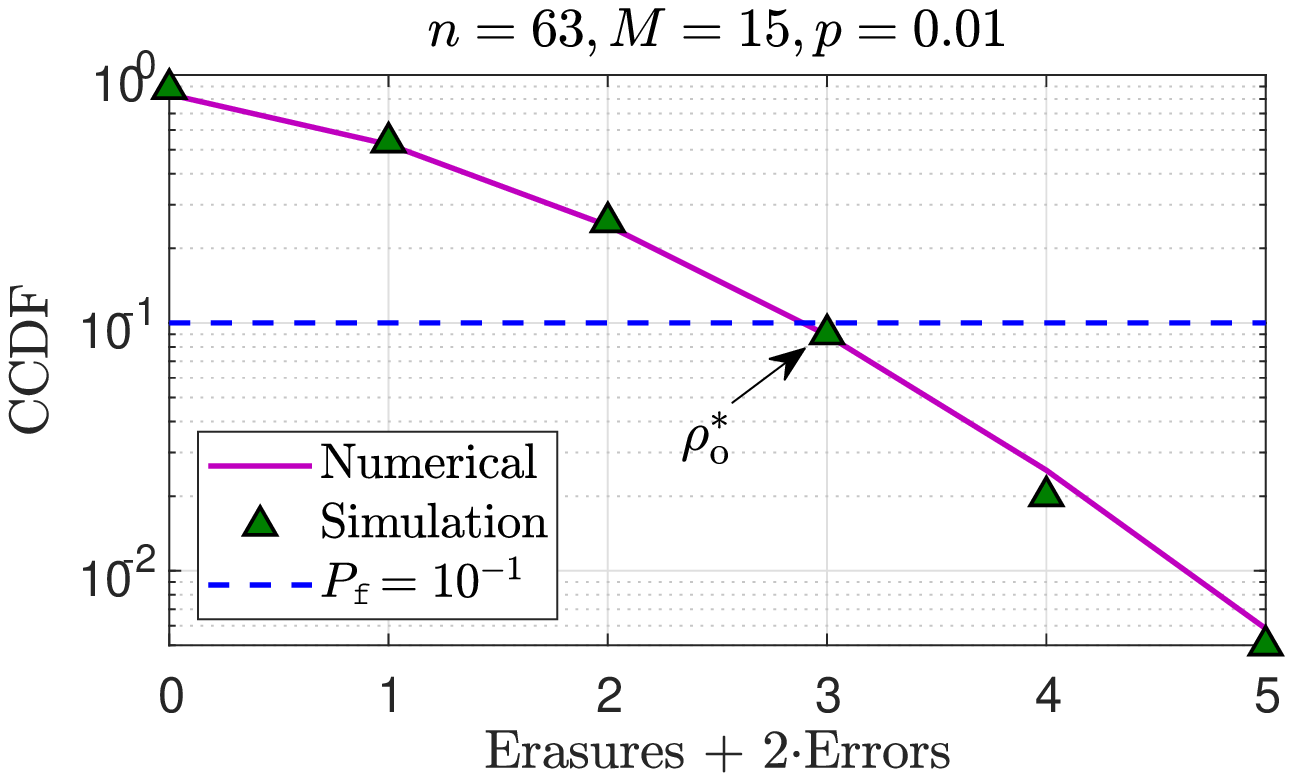}
\end{subfigure}
\caption{\textbf{Left:} Binomial CDFs of erasures and errors, and the resulted trinomial CDF of $W$. \textbf{Right}: Simulation and analytical results of the trinomial CCDF of $W$, along with $\rho_{\mathsf{o}}^{*}$ corresponding to $P_\mathsf{f}=10^{-1}$.}
\label{Fig:OuterChannel}
\end{figure*}

To obtain explicit upper bounds for $\Pers,\Perr$, we substitute into the inequalities of Lemma~\ref{Lem:OuterProbUB} the corresponding values from Lemmas~\ref{Lem:Sucp},~\ref{Lem:ProperMiv}, and~\ref{Lem:MivGreaterThanZero}, together with the relation $\Ppfail=1-\Ppnexc-\Ppmiv$. We can now derive the redundancy of the outer code required to achieve success probability of $P_{\mathsf{s}}$ to recover the batch of $M$ reads.

\vspace*{-0.5ex}
\begin{proposition}
\label{prop:OuterRed}
Let $\cC_{\mathsf{o}}$ be an \textit{MDS} code, and define the random variable $W=\ners+2\nerr$, where $\ners$ and $\nerr$ are the numbers of erasures and errors, respectively. Then,
\begin{align}
	\label{Eq:OuterErr}
	P(W = u) = \hspace*{-1.5ex}\sum_{\underline{m} \in S(M,u)} \frac{M !}{m_0 ! m_1 ! m_2 !} P_\mathsf{cor}^{m_0}\cdot P_{\mathsf{ers}}^{m_1} \cdot P_{\mathsf{err}}^{m_2} ,
\end{align}
for $P_\mathsf{cor} \triangleq 1-P_{\mathsf{ers}}-P_{\mathsf{err}}$ , $\underline{m} = (m_0,m_1,m_2)$,
\begin{equation*}
	S(M,u) \triangleq \left\lbrace \underline{m} \ \left| \begin{matrix} m_1+2m_2=u \\ m_0 +m_1 + m_2 = M
	\end{matrix} \right. \right\rbrace,
\end{equation*}
and the minimal required redundancy $M-k_{\mathsf{o}}$ of $\cC_{\mathsf{o}}$ is given by

\begin{equation}
	\label{Eq:RhoOut}
	\rho_{\mathsf{o}}^* = \min\{w\} \ \ \text{such that} \ \ P\left(W \leq w \right) \geq P_{\mathsf{s}} \ .
\end{equation}
\end{proposition}

\vspace*{-1ex}
\begin{IEEEproof}
The proof is given in Appendix~\ref{Ap:OuterRed}.
\end{IEEEproof}

It is noted that the upper bounds derived in Lemma~\ref{Lem:OuterProbUB} can be used in~\eqref{Eq:OuterErr} instead of the exact probabilities, and provide a sufficient value of $\rho_{\mathsf{o}}^*$, thanks to the probability in~\eqref{Eq:RhoOut} being monotonically decreasing with $\Pers,\Perr$. An optimization procedure of the scheme parameters $\{k_1,\tau,\ell,\mathsf{T}\}$ for obtaining minimal rate is described in~\cite{sps0}.

\subsection{Alignment Performance With SDMS Errors}
\label{subsec:ModAlignmentAnalysis}
To extend the probabilistic analysis to the case of single-deletion multiple-substitutions, we examine in this subsection the alignment success probability under the condition of Definition~\ref{Def:SCmatch}. In the next sub-section we move to examine the inner-decoding outcomes exhibited in Algorithm~\ref{Alg:Cons1ModDec}.

Denote by $\cQpas$,$\cQias$ the probabilities of alignment matches under Definition~\ref{Def:SCmatch} for the proper and improper alignments, respectively.

\vspace*{1ex}
\begin{lemma}
	$	\cQpas \geq \mathsf{F_b}(\ell,p,\mathsf{T}) $, that is, at least as good as in the substitutions-only case.
\end{lemma}
\vspace*{-1ex}
\begin{IEEEproof}
Let $1\leq \isd \leq n+1$ be the index in which the deletion occurred in $\bfy$. Let $\bfz$ be the word obtained from $\bfy$ by a deletion in $\isd$ and no substitutions. By Lemma~\ref{Lem:deloctUB} we have
\begin{align*}
	P(\sdeloct(\bfx,\bfy) \leq \mathsf{T}) \geq P(d_{\mathsf{H}}(f_\ell(\bfx),f_\ell(\bfz)) \leq \mathsf{T}).
\end{align*}
By definition, $\bfx,\bfz$ are related through the substitution channel alone, hence $P(d_{\mathsf{H}}(f_\ell(\bfx),f_\ell(\bfz)) \leq \mathsf{T}) = \mathsf{F_b}(\ell,p,\mathsf{T})$.
\end{IEEEproof} 

For a simplified calculation of the improper-alignment probability, we neglect sub-sequences overlapping with the proper alignment, as for large $L$ these are rare cases. Then, the analysis of Theorem~\ref{Th:RandomDistance} fully captures the misalignment probability, with the threshold $\mathsf{T}$, as stated in the next lemma.

\vspace*{-2ex}
\begin{lemma}
For vectors $\bfy$ that do not overlap with the $\bfY$-indices of $\bfx$, we have
\begin{align*}
	\cQias = P(\delta_\ell \leq \mathsf{T}) ,
\end{align*}
where $\delta_\ell$ is the random variable whose distribution is given in Theorem~\ref{Th:RandomDistance}.
\end{lemma}

For the SDMS model, $\cQpas$ replaces the binomial term $\mathsf{F_b}(\ell,p,\mathsf{T})$ in Lemmas~\ref{Lem:Sucp},~\ref{Lem:ProperMiv}, and $\cQias$ replaces the term $\mathsf{F_b}(\ell,1/2,\mathsf{T})$ in Lemma~\ref{Lem:MivGreaterThanZero}.


\vspace*{-3ex}
\subsection{Inner Decoding With SDMS Errors}
\label{SubSec:ModInnerAnalysis}
For the inner-decoding probabilities under the SDMS model, we bound the random-misvalidation probability (Definition~\ref{Def:MivProb}) for the case of multiple invocations of the inner decoder, as specified in the inner-most loop of Algorithm~\ref{Alg:Cons1ModDec}. To that end, we define $\cPrmiv$ (a modification of $\Prmiv$ in~\eqref{Eq:randMiv}) to be the random-misvalidation probability in the SDMS case. Then we get the following.

\begin{lemma}
The modified random-misvalidation probability $\cPrmiv$ satisfies:
\begin{align*}
	\cPrmiv \leq \frac{V_{n_{\ell}}(t_1)\cdot (n+2)}{2^{n+1-k_2}}.
\end{align*}
\end{lemma}
\begin{IEEEproof}
A necessary condition for misvalidation is that in the single-deletion ball around $\bfy\in\bSig^{n+1}$ there is a word $\bfu$ whose $\bfu_{\cI}$ is in $\cC_2^{[*]}(t_1,\bfv)$ ($*$ represents some particular coset of $\cC_2$; recall definition from~\eqref{Eq:C1MivCandidates}). Equivalently, $\bfy$ is in the single-insertion ball $\mathrm{I}_1(\bfu)$ of at least one vector  $\bfu$ whose $\bfu_{\cI}$ is in $\cC_2^{[*]}(t_1,\bfv)$. Each vector $\bfu_{\cI}\in\cC_2^{[*]}(t_1,\bfv)$ is mapped to $2^{\ell}$ vectors $\bfu$, and each $\bfu$ is mapped to $\left| \mathrm{I}_1(\bfu) \right|=n+2$ $\bfy$-vectors. Upper bounding $\left|\cC_2^{[*]}(t_1,\bfv) \right| $ by $2^{k_2-\ell}\cdot V_{n_{\ell}}(t_1)$, we get that the number of such $\bfy$ vectors is at most\footnote{The actual number may be smaller due to overlaps between the insertion balls.}  $2^{k_2-\ell}\cdot V_{n_{\ell}}(t_1)\cdot2^{\ell}\cdot(n+2)$. Dividing this bound by the number of $\bfy$ vectors $2^{n+1}$, we get the statement.
\end{IEEEproof}
Note that the upper bound on $\cPrmiv$ is roughly factor $(n+2)/2$ larger than $\Prmiv$ in~\eqref{Eq:randMiv}.
Now we can replace $\Prmiv$ with $\cPrmiv$ in Lemmas~\ref{Lem:ProperMiv},~\ref{Lem:MivGreaterThanZero}, and proceed with the analysis as in Sections~\ref{subsec:InnerAnalysis},~\ref{SubSec:OuterChan}
%


\subsection{Numerical Results}

The trinomial CDF of $W$, described in~\eqref{Eq:OuterErr} is demonstrated in the left side of Fig.~\ref{Fig:OuterChannel} (in yellow), compared to the binomial CDFs of erasures (blue dash-dot line) and errors (red dashed line), for $n=M=255$ and for $\Pers=0.1, \Perr=0.01$.

In the right side of Fig.~\ref{Fig:OuterChannel}, the trinomial complementary CDF (CCDF) is shown (in solid purple) for $\Pers,\Perr$ that are evaluated based on the analysis in Sections~\ref{subsec:InnerAnalysis} and \ref{SubSec:OuterChan} for $n=63, k_1=36, \tau=6, \ell=10, M=15, p=0.01$. This is compared with simulation results of encoding and decoding random words, using algebraic codes as described in Section~\ref{SubSec:AlgCons},  while counting the number of erasures and twice the number of errors and evaluating the same CCDF (triangle markers). The required outer redundancy is also marked (horizontal dashed line)  for $P_\mathsf{f}\triangleq 1-P_\mathsf{s} = 10^{-1}$, as the lowest value in x-axis for which the CCDF is lower than $P_\mathsf{f}$.

\begin{figure}[h!]
\centering
\includegraphics[width=0.95\linewidth]{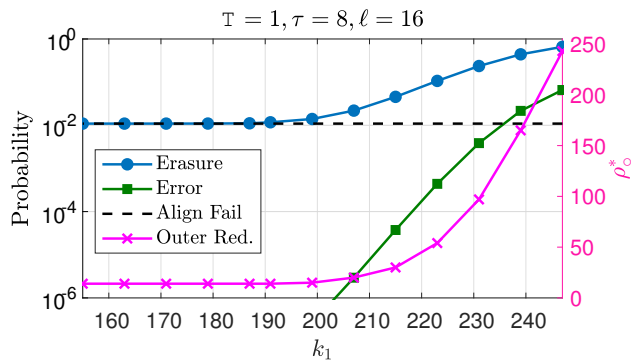}
\caption{Outer channel probabilities and required redundancy as a function of inner-code dimension $k_1$.}
\vspace*{-1ex}
\label{Fig:InnerOuterRed}
\end{figure}

To see the interaction between the inner and outer codes, and to get the idea of their joint optimization, we plot in Fig.~\ref{Fig:InnerOuterRed} some of the interesting variables defined earlier in the section. We first fix the alignment parameters $\mathsf{T}$,$\ell$ and validation parameter $\tau$. Then we plot, as a function of the inner-code dimension $k_1$, the probabilities $P_{\mathsf{ers}}$ (circle markers) and $P_{\mathsf{err}}$ (square markers), for $p=0.01$. Both of these probabilities are monotonically non-decreasing as the inner code gets weaker. We also plot the required outer redundancy~\eqref{Eq:RhoOut} ($\times$ markers, right y-axis) to achieve $P_{\mathsf{s}}>1-10^{-6}$. The code designer is to choose the combination of $k_1=207$ and $\rho_{\mathsf{o}}^*=20$ that minimizes the compression rate, which is $\cR \simeq 0.339$ in this setup. The plot also reveals a floor value for the erasure probability, attributed to failed proper alignments, which cannot be improved by strengthening the inner code.

%

%

\begin{figure}[h!]
\centering
\includegraphics[width=0.95\linewidth]{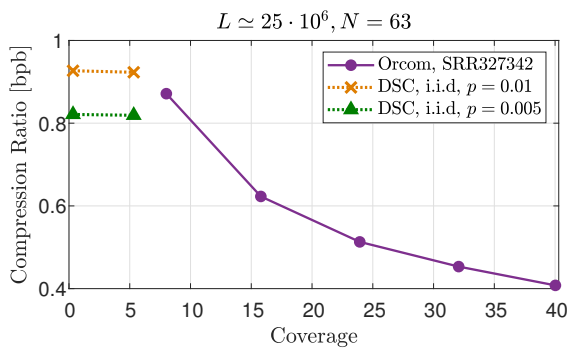}
\caption{Rate comparison between our scheme (DSC) and ORCOM as a fucntion of coverage.}
\vspace*{-1ex}
\label{Fig:RateComp}
\end{figure}

In Fig.~\ref{Fig:RateComp}, we compare the compression rate of the proposed scheme (for $p=0.01,0.005$) to the ubiquitous ORCOM algorithm~\cite{sps34}. For this comparison we use the standard measure of compression ratio in units of bits per base-pair (bpb). Without compression, each $q=4$ base-pair consumes $2$ bits, so the compression rate defined in Proposition~\ref{prop:Rate} is simply the plotted bpb divided by $2$. We use read size of $N = 63$ base pairs for both schemes, which implies an inner-block size of $n=N\log_2(q)=126$ bits in our scheme. For ORCOM, the compression ratios (taken from\cite{sps35}) strongly depend on the sequencing \emph{coverage}~\cite{sps33} (the average number of appearances of a specific symbol from the genome), due to its reliance on internal similarities. Our proposed scheme, in contrast, does not need high coverage, and is shown to achieve comparable rates, even when evaluated on synthetic DNA sequences that have higher entropy. Working at low coverages has significant advantages in complexity and latency.

\begin{figure}[h!]
\centering
\includegraphics[width=0.9\linewidth]{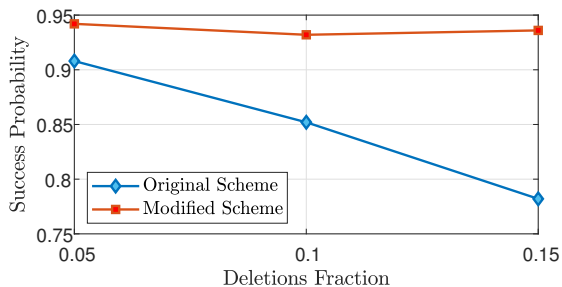}
\caption{Simulation results of alignment+inner-decoding success probability of the proper alignment for the original and modified schemes.}
\label{Fig:InnedDecStat}
\end{figure}

Finally, in Fig.~\ref{Fig:InnedDecStat} we show simulation results of alignment and inner-decoding success probability of the proper alignment for the original (Algorithm~\ref{Alg:Cons1Dec}) and modified (Algorithm~\ref{Alg:Cons1ModDec}) schemes, as a function of the deletion fraction out of the total number of errors. For both curves, we use the same inner code with parameters $[127-\ell,106-\ell,7]$, $\ell=28$, while only modifying the (sub)-candidate selection and the inner-decoding procedures, as described in Algorithm~\ref{Alg:Cons1ModDec}. It is clearly seen that the modified scheme can much better handle deletion errors with significant fractions. 

\section{Further Extensions}
\label{Sec:Future}

\subsection{Multiple Deletions Compensating Distance}
It is possible to extend the shift compensating distance to deal with multiple deletions in the same read.

\vspace*{-2ex}
\begin{definition}
Let $\bfx \in \bSig^n , \bfy \in \bSig^{n+r}$, and let

\vspace*{-3ex}
\begin{align*}
	\underline{t}^* & = \argmax_{\underline{t} \in \mathsf{S}(r,n)} \left\lbrace \sum_{j=0}^{r-1} \Delta\phi_j (\bfx,\bfy; t_j) \right\rbrace , \\
	\mathsf{S}(r,n) & = \left\lbrace \underline{t} =  (t_0, \right. \left. t_1, \dots,t_{r-1}) \in (\mathbb{Z}_0^{+})^r \right| \\
	& \qquad \qquad \left. 1\leq t_0 < t_1 < \dots < t_{r-1} \leq n \right\rbrace .
\end{align*}

The \textbf{shift compensating distance of order $r$} is defined by

\vspace*{-3ex}
\begin{align*}
	\sdelocr (\bfx,\bfy) & \triangleq \phi_r(\bfx,\bfy;n) - \sum_{j=0}^{r-1} \Delta\phi_j (\bfx,\bfy; t_j^*) .
\end{align*}
\end{definition}

It is noted that a greedy approach can be employed, that is for every $1 \leq j \leq n$

\vspace*{-3ex}
\begin{align*}
\tilde{t}^*_j = \argmax_{t : \ t_{j-1} < t \leq n} \left\lbrace \Delta\phi_j (\bfx,\bfy,t) \right\rbrace ,
\end{align*}
where $t_0 = 0$. Under reasonable conditions (e.g. small substitution probability, non-consecutive deletions), it is expected that $\underline{\tilde{t}} = \underline{t}$. Such a greedy approach can be highly advantageous over other metrics since it can be computed in linear time. An exact analysis of the conditions for the greedy approach to match the optimal one is outside the scope of this work.


%
\subsection{Modifying To Single-Insertion}
In this sub-section we briefly describe the slight modifications needed to adapt the single-deletion scheme to a single-insertion scheme. The basis for this adaptation is in the fact that an insertion made in the reference is equivalent to a deletion made in the read. For alignment with insertion in the reference, it would have been natural to reverse the roles of the read and reference in the shift compensating distance, i.e., using $\sdeloc(\bfy,\bfx)$ (where now $|\bfx| > |\bfy|$). Nevertheless, since only the read identifier of $\bfx$ is sent to the decoder for alignment, this is not possible since the the decoder does not have consecutive indices of $\bfx$. Instead, we can define a cumulative Hamming distance with negative shifts, as follows
\begin{align*}
\phi_{-1}(\bfx,\bfy;t) \triangleq \sum_{i=2}^{t+1} x_i \oplus y_{i-1} ,
\end{align*}
cf. Definition~\ref{Def:CumHamDist}. We can now use $\Delta \phi_{-1} (\bfx,\bfy;t) \triangleq \phi_0(\bfx,\bfy;t) - \phi_{-1}(\bfx,\bfy;t)$ to obtain a shift-compensating distance with a single insertion. For inner decoding we use the insertion ball $\mathrm{I}_1(\bar{\bfy}^{(i,j)})$ of every candidate $\bar{\bfy}^{(i,j)} \in \mathrm{Y}$, as a sub-sequence of $\bfY$ of length $n-1$. From this point, the same decoding process as described in Section~\ref{SubSec:InnerMod} is performed.

\vspace*{-0.5ex}

\vspace*{-1ex}
\section{Conclusions}\label{Sec:conclude}
In this paper we addressed the problem of compressing DNA-sequencing reads generated from a genome, while a closely similar reference is available at the decoder side. Specifically, we developed a distributed-source-coding scheme supporting an alignment process at the decoder. As a model for differences between the reference and the reads, we studied both substitutions-only and multiple substitutions with a single deletion (SDMS). For both models we provided decoding algorithms and probabilistic analysis of success probability. For the SDMS model we also introduced a novel alignment metric that is both efficient to evaluate and proven to have good rejection of false alignments.       
As some interesting topics for future research, we suggest the following problems:
\begin{enumerate}
\item Adapting the scheme to optimally suit more complex statistical models of realistic genomic sequences.
\item Developing an efficient algorithm for the calculation of the shift-compensating distance of order $r$, along with its performance analysis.
\item Modifying the scheme to support both deletions and insertions (in the same or different reads).
\end{enumerate}

\vspace*{-1.5ex}
\section*{Acknowledgment}
The authors would like to thank distinguished Prof. Jacob Ziv for valuable discussions and ideas.



%

\bibliographystyle{IEEEtran}
\bibliography{IEEEabrv,References}

\appendices

\section{Proof of Proposition~\ref{prop:OuterRed}}
\label{Ap:OuterRed}
\begin{IEEEproof}
A code with minimum distance $d$ can correct up to $\ners$ erasures and $\nerr$ errors as long as $\ners+2\nerr \leq d-1$ \cite{sps12}. For an MDS code, the minimum distance satisfies $d-1 = \rho_{\mathsf{o}}$. Therefore, whenever $m_1+2m_2 = W \leq \rho_{\mathsf{o}}$ a perfect reconstruction will occur, i.e. $\underline{\hat{\bfa}} = \underline{\bfa}$. As described in Proposition~\ref{prop:outer_success}, we require a probability of at least $P_{\mathsf{s}}$ for that event, i.e. $P(W<\rho_{\mathsf{o}}) \geq P_{\mathsf{s}}$. $S(M,u)$ is the set of triplets $(m_0,m_1,m_2)$ for which their sum equals $M$ and $m_1+2m_2 = u$, and $M!/(m_0!m_1!m_2!)$ enumerates the number of such possible combinations of reads, implying~\eqref{Eq:OuterErr}. Finally, the minimal value of $\rho_{\mathsf{o}}$ satisfying the requirement is $\rho_{\mathsf{o}}^*$.
\end{IEEEproof}

\section{Simple Random Walks Properties}
\label{Ap:RW}
Let $X_i \in \{-1,1\}, i = 1,2,\dots,n$ be an i.i.d sequence of random variables for which $P(X_i=0)=P(X_i=1) = 1/2$. Then, $S_n \triangleq \sum_{i=1}^n X_i$ is called a \emph{simple random walk} in one dimension. We further define $S_0 \triangleq 0$. A specific sequence $(S_1,S_2,\dots,S_n)$ is called a \emph{path}. It is a well known result (e.g.\cite{sps31}) that the probability distribution of $S_n$ is given by $P(S_n = k) = {n \choose (n-k)/2} / 2^n$.
The \emph{reflection principle} is a known property stating that for every path there is a bijection to another path for which the sub-path $(S_{k+1},\dots,S_n)$ is reflected with respect to $S_k$. The next lemma follows.

\begin{lemma}
\label{Lem:RWmax}
Let $M_n = \max_{0\leq i \leq n} \{S_i\}$ be the walk's \emph{maximal value} up to step $n$. Then,
\begin{align*}
	P(M_n = k) = \begin{cases}
		P(S_n = k) & , (n-k) \equiv 0 \mod 2 \\
		P(S_n = k+1) & , (n-k) \equiv 1 \mod 2 .
	\end{cases}
\end{align*}
\end{lemma}

\begin{IEEEproof}
See~\cite{sps32} for detailed proof.
\end{IEEEproof}

\begin{lemma}
\label{Lem:RWjoint}
Let $M_n,T_n$ denote the walk's maximal value and the first step index it was attained up to step $n$, respectively. Then,
\begin{align*}
	P( & M_n = m , T_n = t) =  \\
	& \frac{1}{2} P(M_{n-t} = 0) \cdot \left[ P(S_{t-1} = m-1) - P(S_{t-1} = m+1) \right] 
\end{align*}
\end{lemma}
\begin{IEEEproof}
We have
\begin{align*}
	P(M_n = m, T_n = t) & = P(M_n = m,M_{t-1}=m-1,S_t = m) \\
	& = P(M_n = m | M_{t-1} = m-1, S_t = m) \\
	& \qquad \cdot P(M_{t-1} = m-1, S_t = m).
\end{align*}
Now,
\begin{align*}
	P(M_n = m | & M_{t-1} = m-1, S_t = m) \\
	& = P(M_n = m | S_t = m) = P(M_{n-t} = 0),
\end{align*}
and 
\begin{align*}
	P( M_{t-1} = & m-1, S_t = m) \\
	& = \frac{1}{2} P( M_{t-1} = m-1, S_t = m-1) \\
	& =  \frac{1}{2} \left[ P(M_{t-1} \geq m-1,S_{t-1} = m-1) \right. \\
	& \qquad \left. - P(M_{t-1} \geq m , S_{t-1} = m-1) \right]\\
	& =  \frac{1}{2} \left[P(S_{t-1} = m-1) - P(S_{t-1} = m+1) \right] ,
\end{align*}
where the second term in the last equality is due to the reflection principle. Overall, we get the desired expression.
\end{IEEEproof}

\section{Proof of Theorem~\ref{Th:RandomDistance}: Distribution of the Shift-Compensating Distance}
\label{Ap:docRandom}
Define the random variables $Z_1 \triangleq \phi_1(n)$ and $Z_2 \triangleq \max_{0\leq t\leq n} \{\phi_1(t)-\phi_0(t)\}$. We have
\begin{small}
\begin{align*}
	P(\delta_n = r) = P(Z_1 - Z_2 = r) = \sum_{m=0}^{n-r} P(Z_1=r+m, Z_2 = m) .
\end{align*}
\end{small}
We denote $G_n(\cdot)$ as the number of length-$n$ sequence pairs $\phi_0(i),\phi_1(i)$ subject to the conditions in its arguments. With no conditions, $G_n(\perp)=4^n$. We further denote by $G_n(\cdot|m)$ the number of sequence pairs corresponding to a particular random walk sequence with $M_n=m$, and by $G_n(\cdot|m,t)$ this number when the particular random walk has both $M_n=m$ and $T_n=t$. Additionally, we denote $Q_n(m,t),Q^0_n(m,t)$ as the number of length-$n$ random walk sequences with $M_n=m$ and $T_n=t$, where the former counts walks without and the latter with null steps. A third argument appearing in $Q^0_n$, or to the right of the $|$ in $G_n(\cdot|\cdot)$, represents the number of null steps in the random walk. We can now write
\begin{small}
\begin{align*}
	G_n & (Z_1 = r+m , Z_2 = m) \\ 
	& = G_n(\phi_1(n) = r+m | m)\cdot Q_n^0(m) \\
	& = \sum_{t=m}^{n} G_n(\phi_1(n) = r+m | m,t) Q_n^0(m,t) \\
	& = \sum_{t=m}^{n} \sum_{l=0}^{t-1} G_n(\phi_1(n) = r+m , \phi_0(t) = l | m,t) Q_n^0(m,t) \\
	& = \sum_{t=m}^{n} \sum_{l=0}^{t-1} G_n(\phi_1(n)-\phi_1(t) = r-l , \phi_0(t) = l | m,t) Q_n^0(m,t) . 
\end{align*}
\end{small}

From here, we can split the sequences counting up to and after step $t$, where: (i) the left random walk (up to $t$) has $(m,t)$, (ii) the right random walk (after time $t$, whose value is shifted to start at $0$) has $(m',t')=(0,0)$, and (iii) the function $\phit(k) \triangleq \phi_1(k+t) - \phi_1(t)$ is independent from $\phi_1(i) , 1\leq i \leq t$, and satisfies $\phit(n-t) = r-l$. Hence,
\begin{small}
\begin{align*}
	G&_n (Z_1 = r+m , Z_2 = m) \\ 
	& = \sum_{t=m}^{n} \sum_{l=0}^{t-1} \left[ G_t(\phi_0(t)=l | m,t) Q^0_t(m,t)  \right] \cdot \\
	& \quad \left[ G_{n-t}(\phit(n-t)=r-l|0,0) Q^0_{n-t}(0,0) \right] \\
	& = \sum_{t=m}^{n} \sum_{k=0}^{n-m} \sum_{w=0}^k Q^0_t(m,t,w) Q^0_{n-t} (0,0,k-w) \cdot \\
	&  \sum_{l=0}^{t-1} G_t(\phi_0(t)=l|m,t,w) G_{n-t} (\phit(n-t)=r-l | 0,0,k-w) .
\end{align*}
\end{small}
We now write $Q^0_t (m,t,w) = Q_{t-w} (m,t-w) \cdot {t-1 \choose w}$, since we can eliminate the zeros, evaluate a standard random walk, and multiply by the number of ways to place $w$ zeros along the sequence, excpet for the last step (which must be 1 since it attains the maximum). 

Similarly, $Q^0_{n-t} (0,0,k-w) = Q_{n-t-(k-w)} (0,0) {n-t \choose k-w}$, where in this case there is no requirement on the last step. Hence,

\begin{align*}
G_n & (Z_1 = r+m , Z_2 = m) \\ 
& = \sum_{t=m}^{n} \sum_{k=0}^{n-m} \sum_{w=0}^k \sum_{l=0}^{t-1} G_t(\phi_0(t)=l|m,t,w) Q_{t-w}(m,t-w)  \cdot \\
& \qquad G_{n-t} (\phit(n-t)=r-l | 0,0,k-w ) \\
& \qquad \qquad Q_{n-t-(k-w)}(0,0) \cdot {t-1 \choose w} {n-t \choose k-w} .
\end{align*}

Let $X^{-}_t , X^{+}_t \in \{0,1\}$ be the processes for which $\phi_0(k) = \sum_{i=1}^k X^{-}_t, \phi_1(k) = \sum_{i=1}^k X^{+}_t$. Let $S_t \triangleq \phi_1(t) - \phi_0(t)$, and let $\bar{S}_v$ be obtained from $S_t$ after removing the null steps. Finally, let $\bar{S}^{-}_v , \bar{S}^{+}_v$ be the corresponding (-),(+) walks (derived from $X^{-}_t , X^{+}_t$ in the non-null steps), such that $\bar{S}_v = \bar{S}^{+}_v - \bar{S}^{-}_v$. We now add a condition on the right random walk in its final step, which can take the values $\{-(n-t-(k-w)) , -(n-t-(k-w))+2 , \dots, 0\}$. We also denote $\tilde{S}_k \triangleq S_{k+t}, \check{S}_k \triangleq \bar{S}_{k+t}$.

\begin{align}
\label{Eq:RWcount}
G&_n (Z_1 = r+m , Z_2 = m) \\ 
& = \sum_{t=m}^{n} \sum_{k=0}^{n-m} \sum_{w=0}^k \sum_{l=0}^{t-1} G_t(\phi_0(t)=l|m,t,w) Q_{t-w}(m,t-w)  \cdot \nonumber \\
& \quad \smashoperator[lr]{\sum_{v=0}^{n-t-(k-w)}} G_{n-t} (\phit(n-t)=r-l | 0,0,k-w ,\tilde{S}_{n-t} = -v) \nonumber \\
& \quad Q_{n-t-(k-w)}(0,0,\check{S}_{n-t-(k-w)} = -v) \cdot {t-1 \choose w} {n-t \choose k-w} . \nonumber
\end{align}

We can now enumerate the different terms in the sum above. 
\begin{align*}
Q&_{t-w}(m,t-w) = 2^{t-w} \cdot P(M_{t-w}=m,T_{t-w}=t) \\
& = \left[{t-w-1 \choose (t-w+m)/2-1} - {t-w-1 \choose (t-w+m)/2} \right] \\
& = \left[{t-w-1 \choose (t-w-m)/2} - {t-w-1 \choose (t-w-m)/2-1} \right]\\
& = \frac{m}{t-w} {t-w \choose (t-w-m)/2} ,
\end{align*}
where the first equality is obtained from Lemmas~\ref{Lem:RWmax},~\ref{Lem:RWjoint}.

For each step $i$ for which $X^{+}_i - X^{-}_i = 0$, we separate between a \emph{type-I zero} for which $X^{+}_i = X^{-}_i = 0$ (no increments to $\phi_0,\phi_1$ in this step), and a \emph{type-II zero} for which $X^{+}_i = X^{-}_i = 1$ (both $\phi_0,\phi_1$ increase by 1). For the non-null steps up to $t$ we have $\bar{S}^{+}_{t-w} + \bar{S}^{-}_{t-w} = t-w$ and $\bar{S}^{+}_{t-w} - \bar{S}^{-}_{t-w} = m$, hence $\bar{S}^{+}_{t-w} = (t-w+m)/2$ and $\bar{S}^{-}_{t-w} = \alpha \triangleq (t-w-m)/2$. Moreover, since $l$ determines the number of steps up to $t$ for which $X^{+}_i - X^{-}_i \in \{0,-1\}$, there must be $\beta \triangleq l-\alpha$ type-II zeros (since there are $\alpha$ negative steps), and therefore $w-\beta$ type-I zeros up to step $t$. Now, $G_t({\phi_{0}(t) = l | m,t,w})$ is obtained from counting the possible ways to place $\beta$ type-II zeros and $w-\beta$ type-I zeros from the $w$ zeros, i.e. ${w \choose \beta}$. For the non-null steps after $t$, we use the following lemma.

\begin{lemma}
For a non-positive simple symmetric random walk, the number of possible sequences with $a$ negative steps and $b\leq a$ positive steps is
\begin{align*}
	B(a,b) = {a+b \choose b} - {a+b \choose b-1} = \frac{a-b+1}{a+b+1} {a+b+1 \choose b} .
\end{align*}
\end{lemma}
\begin{IEEEproof}
The proof is based on similar considerations as in the proof of Theorem 1 in~\cite{sps36}, with a reversal of roles between positive and negative steps.
\end{IEEEproof}

Based on this lemma, we get
\begin{align*}
Q&_{n-t-(k-w)}(0,0,\bar{S}^{'}_{n-t-(k-w)} = -v )\\
& = B\left(\frac{n-t-(k-w)+v}{2},\frac{n-t-(k-w)-v}{2}\right) \\
& = \frac{v+1}{n-t-(k-w)+1} {n-t-(k-w)+1 \choose (n-t-(k-w)-v)/2} .
\end{align*}
Using similar considerations as above, we get that there are $\eta \triangleq r-l-\gamma$ type-II zeros and $k-w-\eta$ type-I zeros after step $t$, where $\gamma \triangleq (n-t-(k-w)-v)/2$. Again, by counting the number of possible placements of zeros we get that $G_{n-t} (\phit(n-t)=r-l | 0,0,k-w ,\tilde{S}_{n-t} = -v) = {k-w \choose \eta}$. 

By substituting all of the above into~\eqref{Eq:RWcount}, collecting binomial terms to multinomial terms and summing over $m$ from $1$ to $n$, we complete the proof.



\ifCLASSOPTIONcaptionsoff
\newpage
\fi

\end{document}